\documentclass[12pt,preprint]{aastex}

\shorttitle{MIRI surveys}
\shortauthors{Bonato et al.}

\usepackage{graphicx}
\usepackage{natbib}
\usepackage{rotating}
\usepackage{lscape}
\usepackage{url}
\usepackage{amssymb}
\usepackage{morefloats}
\usepackage[breaklinks=true]{hyperref}
\usepackage{breakcites}

\usepackage[normalem]{ulem} 
\usepackage{soul} 

\makeatletter
\newcommand\footnoteref[1]{\protected@xdef\@thefnmark{\ref{#1}}\@footnotemark}
\makeatother

\begin{document}






\title{Exploring the evolution of star formation and dwarf galaxy properties with JWST/MIRI serendipitous spectroscopic surveys}

\author{Matteo Bonato\altaffilmark{1}, Anna Sajina\altaffilmark{1}, Gianfranco De Zotti\altaffilmark{2,3}, Jed McKinney\altaffilmark{1}, Ivano Baronchelli\altaffilmark{4}, Mattia Negrello\altaffilmark{5}, Danilo Marchesini\altaffilmark{1}, Eric Roebuck\altaffilmark{1}, Heath Shipley\altaffilmark{1}, Noah Kurinsky\altaffilmark{6}, Alexandra Pope\altaffilmark{7}, Alberto Noriega-Crespo\altaffilmark{8}, Lin Yan\altaffilmark{4} and Allison Kirkpatrick\altaffilmark{9} }

\altaffiltext{1}{Department of Physics \& Astronomy, Tufts University, 574 Boston Avenue, Medford, MA}
\altaffiltext{2}{INAF, Osservatorio Astronomico di Padova, Vicolo Osservatorio 5, I-35122 Padova, Italy}
\altaffiltext{3}{SISSA, Via Bonomea 265, I-34136 Trieste, Italy}
\altaffiltext{4}{California Institute of Technology, Pasadena, CA}
\altaffiltext{5}{School of Physics and Astronomy, Cardiff University, Queen’s Buildings, The Parade, Cardiff CF24 3AA, UK}
\altaffiltext{6}{Department of Physics, Stanford University, Stanford, CA}
\altaffiltext{7}{Department of Astronomy, University of Massachusetts Amherst, Amherst, MA}
\altaffiltext{8}{Space Telescope Science Institute, 3700 San Martin Dr., Baltimore, MD}
\altaffiltext{9}{Department of Astronomy, Yale University, New Haven, CT}







\begin{abstract}
The James Webb Space Telescope's  Medium Resolution Spectrometer (MRS), will offer nearly 2 orders of magnitude improvement in sensitivity and $>$\,3\,$\times$ improvement in spectral resolution over our previous space-based mid-IR spectrometer, the {\sl Spitzer} IRS. In this paper, we make predictions for spectroscopic pointed observations and serendipitous detections with the MRS. Specifically, pointed observations of \textit{Herschel} sources require only a few minutes on source integration for detections of several star-forming and active galactic nucleus lines, out to z$=$3 and beyond. But the same data will also include tens of serendipitous 0$\lesssim$z$\lesssim$4 galaxies per field with infrared luminosities ranging $\sim10^6-10^{13}$\,L$_{\sun}$. In particular, for the first time and for free we will be able to explore the  $L_{\rm IR}<10^{9}\,L_{\sun}$ regime out to $z\sim3$. We estimate that with $\sim$\,100 such fields, statistics of these detections will be sufficient to constrain the evolution of the low-$L$ end of the infrared luminosity function, and hence the star formation rate function. The above conclusions hold for a wide range in potential low-$L$ end of the IR luminosity function, and accounting for the PAH deficit in low-$L$, low-metallicity galaxies.

\end{abstract}


\section{Introduction}\label{sect:intro}

Over the last couple of decades it has become clear that the bulk of the galaxy star formation (SF) and supermassive black hole (SMBH) accretion in the Universe occurred in the redshift interval 1$\lesssim$z$\lesssim$3 (\citealt{Merloni2008}; \citealt{Madau14}; \citealt{Delvecchio2014}). The correlation between the mass of SMBHs, located at the galaxy centers, and the properties of the spheroidal stellar components \citep[see][for reviews]{FerrareseFord2005,KormendyHo2013} reveal a connection between the black hole growth and the build up of the mass in stars (see e.g. \citealt{Hopkins2006,Hopkins2008}). The debate about the nature of this connection, i.e. a direct interaction, an indirect connection, or a correlation arising from availability of free baryons, is open (see e.g. \citealt{Peng2007}; \citealt{Jahnke2011}; \citealt{Graham2013}). Moreover, the evolution of this connection with cosmic time is largely unknown. In order to shed light on this issue, a detailed investigation of the physical processes operating in large samples of cosmic sources is necessary. Moreover, the most active SF phases of galaxies and the associated active galactic nucleus (AGN) growth are severely dust-obscured (see e.g. \citealt{Burgarella13}; \citealt{Madau14}) and are therefore best studied at IR wavelengths. 

In particular, the mid-IR (MIR) regime (here defined as rest-frame $\sim$2-30$\mu$m) is rich in spectral lines and features that provide excellent diagnostics of the power source (SF or AGN) as well as of the level of obscuration and dust composition \citep[e.g.][]{yan07,Sajina2007,pope08,dasyra09,Sajina2009,Sajina2012,Kirkpatrick2012,Kirkpatrick2015}. In this regime, fine-structure lines come either from star forming regions or from nuclear activity or from both. MIR line ratio diagrams can be used to identify composite sources and to distinguish between emission from star forming regions and emission excited by nuclear activity (\citealt{Genzel1998}; \citealt{Lutz1999}; \citealt{Sturm02}; \citealt{Armus2007}; \citealt{Farrah2007}; ; \citealt{Ho2007}, \citealt{Yuan2010}). The PAH luminosities have been calibrated as star formation rate (SFR) indicators (e.g. \citealt{Roussel2001}, \citealt{ForsterSchreiber2004}, \citealt{Peeters2004}, \citealt{Desai2007}; \citealt{Shipley2016}). 

Thanks to its unprecedented sensitivity and resolution (\citealt{Glasse2015}), a giant leap in this field is expected from the James Webb Space Telescope (JWST\footnote{\url{http://www.jwst.nasa.gov}}; \citealt{Gardner2006}, \citealt{Windhorst2009}, \citealt{Finkelstein2015}) Mid-Infrared Instrument (MIRI; \citealt{Rieke2015}). For example its Medium Resolution Spectrometer (MRS) will provide a gain in sensitivity of 1-2 orders of magnitudes over the {\sl Spitzer} Infrared Spectrograph (\citealt{Houck04}) and $\gtrsim$3\,$\times$ gain in spectral resolution. This advancement will allow us to obtain, at high-$z$, the high-/medium-resolution spectra necessary for the detections of the fine-structure AGN lines, whose study was mainly limited to the local Universe so far (e.g. \citealt{Sturm02}; \citealt{Armus2007}; \citealt{Farrah2007}; \citealt{Dasyra2011}; \citealt{Inami2013}). Therefore, MRS pointed observations will make it possible to not only obtain spectroscopic redshifts for essentially any known dusty galaxy population across the bulk of cosmic time, but also to accurately measure the contributions from stellar and AGN activity to their luminosity. 
 

Current photometric surveys of dusty galaxies are biased toward higher luminosity sources since even in the local Universe they are limited to L$_{\rm IR}>$4$\times$10$^{7}$L$_{\sun}$ galaxies and by $z$\,$\sim$\,2, the peak epoch for SF and black hole accretion activity, they are limited to L$_{\rm IR}>$3$\times$10$^{11}$L$_{\sun}$ galaxies (see e.g. \citealt{Sanders2003}, \citealt{Eales2010}, \citealt{Clemens2013}, \citealt{Magnelli11}, \citealt{Gruppioni2013}, \citealt{Lutz2014})\footnote{In \citet{Sanders2003}, the minimum L$_{\rm IR}$ in the Revised Bright Galaxy Sample (mean redshift of the sample z$\sim$0.01), observed by the Infrared Astronomical Satellite (96\% of the sky) at 12, 25, 60 and 100$\mu m$ is $\sim$4$\times$10$^{7}L_{\sun}$ (only 3 galaxies of the sample have L$_{\rm IR}<$10$^{8}$L$_{\sun}$). At z$\sim$2, \citet[in the redshift bin $1.8\lesssim z\lesssim2.3$]{Magnelli11} and \citet[$1.7\lesssim z\lesssim2.0$]{Gruppioni2013} achieve a minimum IR luminosity of $\sim$3$\times$10$^{11}L_{\sun}$. \citet{Magnelli11} used \textit{Spitzer} MIPS (24, 70$\mu m$) and IRAC (3.6$\mu m$) observations of the GOODS North and South fields (area $\sim$100\,arcmin$^{2}$). \citet{Gruppioni2013} used \textit{Herschel} PACS observations (at 70, 100 and 160$\mu m$) in combination with \textit{Herschel} SPIRE data (at 250, 350 and 500$\mu m$), over the GOODS North and South, Extended Chandra Deep
Field South and Cosmic Evolution Survey areas (total area $\sim$2.4\,deg$^{2}$).}. The minimum IR luminosities achieved by the deepest \textit{Spitzer} MIR  spectroscopic observations are $\sim$10$^{9}L_{\sun}$ at z$\sim$0 and $\sim$10$^{11}L_{\sun}$ at z$\sim$2 (see \citealt{ODowd2009}, \citealt{dasyra09}, \citealt{Shipley2013}, \citealt{Sajina2008}). The unprecedented sensitivity of the MRS will allow us to access IR luminosities about two orders of magnitude lower than current limits. The space density of such dwarf galaxies is high enough that we expect to be able to serendipitously detect them within the fields of any MRS pointed observation. Therefore, not only will the JWST-MRS allow us to gain much greater insight into the physical conditions and power sources of known dusty galaxy populations across cosmic history, but it will also allow for the discovery of new populations in the hitherto largely unknown low-$L$, low-metallicity, high-$z$ regime. 

Indeed, the $L_{\rm{IR}}\lesssim 10^9$\,L$_{\odot}$ luminosity regime is largely unexplored and poorly constrained by current galaxy evolution models. So far, IR continuum and PAH emission of dwarf star-forming galaxies has been studied only in the Local Group and in other nearby sources (\citealt{Gallagher1991}; \citealt{Israel1996}; \citealt{Contursi1998,Contursi2000}; \citealt{Sturm2000}; \citealt{Reach2000}; \citealt{Vermeij2002}, \citealt{Hunter2001}; \citealt{Houck2004}; \citealt{Engelbracht2005,Engelbracht2008}; \citealt{Jackson2006}, \citealt{Madden2006,Madden2013}, \citealt{Galametz2009}, \citealt{Remy-Ruyer2013}, \citealt{Fisher2014}). These studies suggest that the SEDs of low-metallicity dwarf galaxies may be very different than typical star-forming galaxies (see e.g. \citealt{Houck2004}). The density of these low-$L$ star-forming dwarf galaxies in the Universe is unknown (see e.g. \citealt{Lutz2014}). Additionally, the relationships between continuum and line luminosity in this luminosity range are very unsure. This is due to the fact that the physical conditions in such low-$L$ galaxies differ from those in their high-L counterparts, which affects line excitation. Therefore a simple extrapolation, at low luminosities, of the line-to-continuum relations calibrated at high-L could be deceiving. For example, the well-known mass-metallicity relation (\citealt{Tremonti2004}; \citealt{Zahid2013}) indicates that the metallicity decreases at lower masses (and therefore at lower luminosities). This effect should be stronger at higher redshifts, since metallicity decreases with increasing $z$ (\citealt{Zahid2013}). MRS detections (but also non-detections) of these low-$L$ galaxies will allow us to explore the physical properties of such sources.

In this paper, we look at the star-forming and AGN line detectability for MIRI MRS pointed observations. We present predictions for the number, luminosity and redshift distribution of serendipitously detected galaxies and AGN within the fields of the same pointed observations. 
We make our predictions based on three different assumptions about the unknown low-$L$ end of the IR luminosity function, and accounting for the potential PAH deficit in low-$L$, low-metallicity galaxies.

The paper is organized as follows. In Section\,\ref{sect:miri}, we present the key technical aspects of the MRS relevant to this paper. In Section\,\ref{sect:mod_approach}, we summarize our modelling approach. In Section\,\ref{sect:MRS_surveys}, we present predictions for MRS pointed observations and serendipitous surveys.
In Section\,\ref{sec:discussion}, we discuss the implications of our results for the study of the galaxy-AGN (co-)evolution and of the low-luminosity galaxy properties.
Section\,\ref{sect:concl}, contains a summary of our main conclusions. Throughout this paper we adopt a flat $\Lambda \rm CDM$ cosmology with matter density $\Omega_{\rm m}\sim0.31$, dark energy density $\Omega_{\Lambda}\sim0.69$ and Hubble constant $h=H_0/100\, \rm km\,s^{-1}\,Mpc^{-1}\sim0.67$ \citep{Planck2015}.

\section{MRS overview} \label{sect:miri}

The MIRI MRS instrument (\citealt{Wells2015}) covers wavelengths from $\sim$4.9 to $\sim$28.8$\,\mu$m\footnote{The full 4.9-28.8$\,\mu$m wavelength coverage needs 3 different grating angle settings. This fact must be taken into account in the computation of the overheads, since the full spectra can be obtained in a time \,3\,$\times$ the on-target integration time used throughout this paper.}, with R$\sim$2200-3500 and Field of View (FoV) from $\sim$3.0''$\times$3.9'' to $\sim$6.7''$\times$7.7''\footnote{\label{note}\url{http://ircamera.as.arizona.edu/MIRI/performance.htm}}. It consists of four Integral Field Units, working simultaneously (\citealt{Gordon2015}). Consistent with \citet{Glasse2015}, the reference extended source
detection limits\footnoteref{note} (10$\sigma$, 10,000s) for the four channels are 1.8$\times$10$^{-20}$W m$^{-2}$/arcsec$^{2}$ (4.87$\lesssim\lambda\lesssim$7.76$\mu$m; R$\sim$3500; FoV$\sim$3.0''$\times$3.9''), 9.0$\times$10$^{-21}$W m$^{-2}$/arcsec$^{2}$ (7.45$\lesssim\lambda\lesssim$11.87$\mu$m; R$\sim$2800; FoV$\sim$3.5''$\times$4.4''), 5.0$\times$10$^{-21}$W m$^{-2}$/arcsec$^{2}$ (11.47$\lesssim\lambda\lesssim$18.24$\mu$m; R$\sim$2700; FoV$\sim$5.2''$\times$6.2'') and 1.9$\times$10$^{-20}$W m$^{-2}$/arcsec$^{2}$ (17.54$\lesssim\lambda\lesssim$28.82$\mu$m; R$\sim$2200; FoV$\sim$6.7''$\times$7.7''). 

The spatial resolution varies from $\sim$0.2 arcsec at 4.9$\mu$m to $\sim$1.1 arcsec at 28.8$\mu$m. Serendipitous detections within the fields of the pointed observations\footnote{Hereafter, by number of serendipitous detections ``per FoV'' we mean the total number of detections (at 5$\sigma$ level) per single exposure, inside the different (superimposed) FoVs of the 4 MRS channels. The maximum area of sky covered in a single exposure is given by the FoV of the highest-wavelength channel (see \citealt{Wells2015}).} are operationally supported only for the MRS, not for the Low Resolution Spectrometer (\citealt{Kendrew2015}). 

\section{Modeling approach}\label{sect:mod_approach}

\subsection{The evolution of the IR luminosity function}\label{subsect:evol}

\begin{figure}  
\centering
    \includegraphics[trim=2.4cm 2.5cm 3.1cm
    3.1cm,clip=true,width=0.99\textwidth, angle=0]{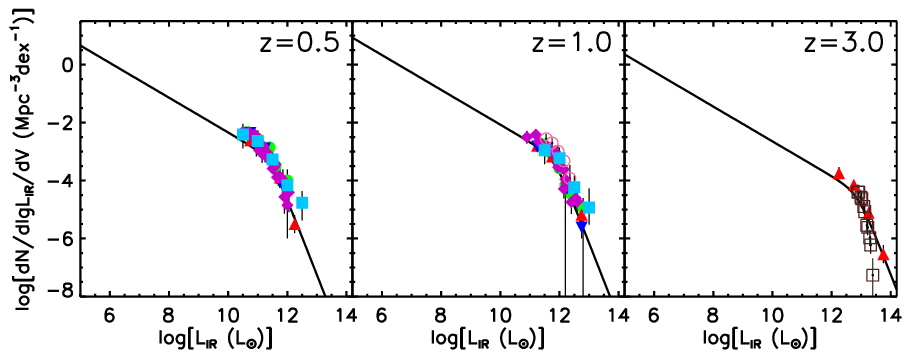}
  \caption{Total IR LFs, at z\,=\,0.5, 1.0 and 3.0, derived from the \citet{Kurinsky2016} model (solid black lines), compared to the following observational estimations taken from literature: \citet[filled cyan squares]{LF05}, \citet[open pink circles]{Caputi07}, \citet[filled violet diamonds]{Magnelli09}, \citet[filled downward blue triangles]{Rodighiero10}, \citet[open brown squares]{Lapi11}, \citet[filled green circles]{Magnelli13} and \citet[filled upward red triangles]{Gruppioni2013}.}
  \label{fig:lf_compare}
\end{figure}

Our reference model for the evolution of the IR luminosity function is the model used in \citet{Kurinsky2016}. The functional form for the luminosity function is a double power-law (e.g. \citealt{Negrello2013}): $\Phi_{PL} = \Phi^*\left[{\left(\frac{L}{L^*}\right)}^{\alpha}+\left(\frac{L}{L^*}\right)^{\beta}\right]^{-1}$; where $\Phi^*(z) = \Phi^*_0(1+z)^p$ and $L^*(z) = L^*_0(1+z)^q$. $p$ denotes density evolution and $q$ denotes luminosity evolution; at two break redshifts the power of the density ($z_{bp}$) and luminosity ($z_{bq}$) evolution changes. The evolution of the AGN fraction is given by a power-law of the form: $f_{AGN,z}=f_{AGN,0}*[log(L_{\rm{TIR}})/12]^{12}*(1+z)^t$; where $f_{AGN,0}$ represents the $z\sim0$ AGN fraction at $log(L_{\rm{TIR}}/L_{\sun})$=12. $t\equiv t_{1}$ until a break redshift, $z_{bt}$, and $t\equiv t_{2}$ at higher z. The fraction of composite sources in respect to all AGN is given by a single parameter: $f_{comp}$. The adopted best-fit values of the parameters (taken from \citealt{Kurinsky2016}) are as follows: $\log(L_0^*/L_\odot)$=$ 10.86^{+  0.06}_{-  0.07}$; $\log(\Phi^*_0)$=$ -3.24^{+  0.06}_{-  0.05}$; $z_{bp}$=$1.66^{+1.43}_{-0.00}$; $z_{bq}$=$  1.85^{+  1.20}_{-0.29}$; $p_{1}$=$0.16^{+1.54}_{-1.53}$; $q_{1}$=$3.33^{+0.89}_{-0.86}$; $p_{2}$=$-3.32^{+2.33}_{-1.91}$; $f_{AGN,0}$=$  0.26^{+0.13}_{-0.10}$; $t_{1}$=$-0.14^{+0.87}_{-0.67}$; $t_{2}$=$0.12^{+4.38}_{-3.25}$; $z_{bt}$=$2.60^{+0.63}_{-0.44}$; $f_{comp}$=$0.33^{+0.43}_{-0.10}$. In such a model the slope parameters are set: $\alpha=2.6$; $\beta=0.6$. These best-fit parameters are found by fitting HerMES-COSMOS MIPS/SPIRE color-magnitude diagrams. This data set yields strong constraints on the evolution of the IR LF up to $z \sim 1.5$. On the other hand, it is not deep enough to effectively constrain the model at higher redshifts. We are working on extending these constraints to higher redshifts by applying the code to deeper samples as well (Bonato et al. 2017 in prep.). 
For the purposes of this paper we have further required consistency of the model with observational determinations of the IR LF at higher z. Observational estimates of the IR LF are currently available up to $z\simeq 4$ (\citealt{Gruppioni2013}). This additional requirement implied the modification of only one parameter, $q_2$. We adopt here $q_{2}=1.6$ to fit literature LF determinations at $z\gtrsim2$. Figure\,\ref{fig:lf_compare} compares our reference model LF at three representative redshifts (0.5, 1 and 3) with the available data. 

The adopted SED library, based on \citet{Kirkpatrick2015}, includes templates for star-forming galaxies, composites, and AGN as well as their redshift and luminosity evolution. It is based on a sample of 343 dusty galaxies with mid-IR IRS spectra as well as a wealth of ancillary data covering the near- to far-IR regime. In particular, the IRS spectra allow for accurate spectral classification into the three different SED types. This library does not include a low metallicity starburst template.

In our reference model, as in all existing IR luminosity function models, the low-$L$ regime is pure extrapolation. Even in the local Universe we have constraints on the IR LF only essentially above $\sim$10$^{8}$L$_{\sun}$; see e.g. \citet{Sanders2003}, \citet{Clemens2013}, \citet{Lutz2014}. Indeed, assuming a constant, relatively steep, $<L_*$ slope quickly reaches unrealistically large space densities: the maximum values found by deep optical surveys are $\sim 0.1\,\hbox{Mpc}^{-3}$dex$^{-1}$ \citep[e.g.,][]{Driver2012}. Generally, we expect a break of the IR LFs at low luminosities for a simple physical reason: very small galaxies have a low binding energy and therefore cannot bear substantial SFR because the mechanical energy released by only a few supernovae is sufficient to unbind the gas, thus halting the SF. At low-$z$, low IR luminosities (typically below $\sim 10^{9}\hbox{--}10^{10} L_\odot$) may be due to dust heated not by newly born stars but by the general radiation field produced by older stellar populations. Obviously in this case we do not expect lines excited by SF. Note also that a fraction of  low IR luminosity objects are massive early-type galaxies, with high bolometric luminosity but almost devoid of gas and dust and consequently with very low SFR.

\begin{figure}  
\centering
    \includegraphics[trim=5.5cm 1.0cm 2.5cm
    1.5cm,clip=true,width=0.49\textwidth, angle=0]{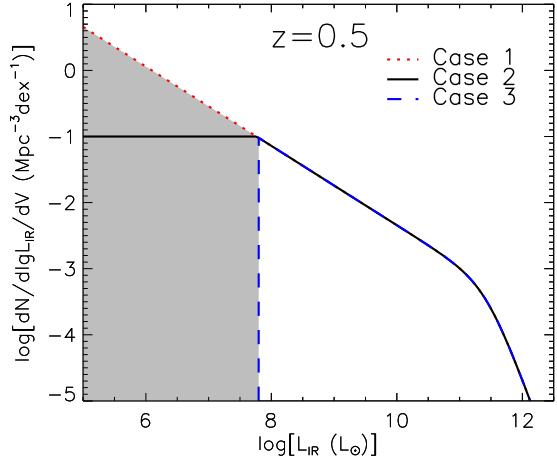}
  \caption{At $z$\,=\,0.5, the total IR LF of the \citet{Kurinsky2016} model with three different assumptions for the low-$L$ end (shaded region): case 1 - extrapolate current models down to 10$^{5}$L$_{\sun}$ (dotted red line); case 2 - break such that all low-$L$ galaxies have a density of 0.1\,Mpc$^{-3}$dex$^{-1}$ (solid black line); case 3 - growth until a density of 0.1\,Mpc$^{-3}$dex$^{-1}$ and null after that (dashed blue line).}
  \label{fig:LF_models2}
\end{figure}

To account for the above uncertainties, we test different assumptions about the low-luminosity end of the IR LF. Specifically, we test three cases (shown in Fig.~\ref{fig:LF_models2}, for a reference redshift of $z$=0.5): 1) extrapolate the current model down to 10$^{5}$L$_{\sun}$ (upper limit); 2) level off the faint end of the IR LF at a density of 0.1\,Mpc$^{-3}$dex$^{-1}$ (intermediate case); 3) impose a low-luminosity cut-off, at all $z$,  at the luminosity corresponding to a density of 0.1\,Mpc$^{-3}$dex$^{-1}$ (lower limit). The last case is for the extreme scenario where at low-$L$ we do not expect any SF or AGN lines (what we ultimately detect with the MRS) as 100\% of the luminosity is due to old stellar populations. These three cases are all compatible with the current constraints given by the observational determinations of the Cosmic Infrared Background (CIB) intensity. In fact the contribution of these very low-L sources to the total CIB is minor, the variation is only a few percents among the three cases. In this paper the ``intermediate case'' for the \citet{Kurinsky2016} model is adopted as reference model, but we compare the results with those obtained using the two extreme cases on either side of the reference model (Case 1 \& Case 3). 

\subsection{Galaxy sizes}\label{subsect:size}

Because of the very small FWHM, a significant fraction of the galaxies will be spatially resolved by the JWST. In order to estimate the galaxy sizes (depending mainly on luminosity and stellar mass) in our simulations, we derived their SFR using the $L_{\rm IR}$/SFR relation by \citet{Clemens2013}.
The SFR was converted into stellar mass using the main sequence empirical correlations of star-forming galaxies obtained by \citet{Kurczynski2016} in the redshift range 0.5$\lesssim$z$\lesssim$3.0. For z$<$0.5, we used the \citet{Peng2010} correlation, measured at z$\sim$0.2; at higher z, the relations (at z$\sim$4, 5 and 6) are derived by \citet{Salmon2015}. Finally the stellar mass content was converted into effective radius through the \citet{Shen2003} best-fit relation\footnote{their eq.18, for late-type galaxies, with parameter values listed in the last line of their Tab.1; we considered a range of effective radii (1-10Kpc) consistent with their fitting (see their Fig.11).} at $z<1$ and through the relation estimated by \citet{Bruce2012}\footnote{For their sample of $1<z<3$ massive galaxies they find a median size of a factor $\sim$2.3 times smaller than comparably massive local galaxies.} at higher redshift. For example, for a galaxy having $L_{\rm IR}\sim10^{12}\,L_{\sun}$ we obtain an effective radius of $\sim$1.13'' (at z$\sim$0.5; corresponding to $\sim$6.6\,kpc) and $\sim$0.14'' (at z$\sim$2.0; $\sim$1.1\,kpc).

We caution that the sizes associated with some simulated galaxies in our analysis are probably overestimated. Specifically, the SFR/M$_{*}$ ratio for galaxies that are starbursting could be significantly higher than the main sequence relations used here (see e.g \citealt{Rodighiero2014}). Moreover, there is evidence that the sizes of galaxies in the MIR may be smaller than in the optical regime (see e.g. \citealt{Diaz-Santos2010}, \citealt{Xu2014}).  Given that more compact galaxies of the same luminosity are easier to detect than more extended ones, this potential overestimation of the sizes means that our results can be viewed as a conservative lower limit on the numbers of serendipitous detections. 

\subsection{Correlations between continuum and line luminosity}\label{subsect:line_vs_IR}

To estimate the counts of galaxy and AGN line detections by MIRI surveys, we coupled the redshift dependent IR luminosity functions of the source populations with relationships between line and continuum luminosities. 

We used the calibrations derived by \citet{Bonato2014a,Bonato2014b,Bonato2015} for 33 MIR fine-structure lines (placed in the rest-frame wavelength range 3.03$\lesssim \lambda \lesssim$25.98$\mu$m), 4 PAH lines at 6.2, 7.7, 8.6 and 11.3$\mu$m and two silicate bands (in emission and in absorption) at 9.7$\mu$m and 18.0$\mu$m. 

The 33 MIR fine-structure lines are the following:
\begin{itemize}
\item 3 coronal region lines: [MgVIII]\,3.03, [SiIX]\,3.92 and [SiVII]\,6.50$\,\mu$m;
\item 13 AGN fine-structure emission lines: [CaIV]\,3.21, [CaV]\,4.20, [MgIV]\,4.49, [ArVI]\,4.52, [MgV]\,5.60, [NeVI]\,7.65, [ArV]\,7.90, [CaV]\,11.48, [ArV]\,13.09, [MgV]\,13.50, [NeV]\,14.32, [NeV]\,24.31 and [OIV]\,25.89$\,\mu$m;
\item 9 stellar/HII region lines: [ArII]\,6.98, [ArIII]\,8.99, [SIV]\,10.49, HI\,12.37, [NeII]\,12.81, [ClII]\,14.38, [NeIII]\,15.55, [SIII]\,18.71 and [ArIII]\,21.82$\,\mu$m;
\item 3 lines from photodissociation regions: [FeII]\,17.93, [FeIII]\,22.90 and [FeII]\,25.98$\,\mu$m;
\item 5 molecular hydrogen lines: H$_{2}$\,5.51, H$_{2}$\,6.91, H$_{2}$\,9.66, H$_{2}$\,12.28 and H$_{2}$\,17.03\,$\mu$m.
\end{itemize}

For the PAH\,\-6.2, PAH\,\-7.7, PAH\,\-8.6, PAH\,\-11.3, Sil\-9.7, Sil\-18.0, H$_{2}$\,\-9.66, [SIV]\-10.49, H$_{2}$\,\-12.28, [NeII]\-12.81, [NeV]\-14.32, [NeIII]\-15.55, H$_{2}$\,\-17.03, [SIII]\-18.71, [NeV]\-24.31 and [OIV]\-25.89$\,\mu$m lines, we used the relationships derived by \citet{Bonato2014a,Bonato2014b,Bonato2015} on the basis of observations collected from the literature. For all the other lines, with either insufficient or missing data, the line to continuum luminosity relations were derived in \citet{Bonato2015} using the IDL Tool for Emission-line Ratio Analysis (ITERA)\footnote{http://home.strw.leidenuniv.nl/~brent/itera.html} written by Brent Groves. ITERA makes use of the library of published photoionization and shock models for line emission of astrophysical plasmas produced by the Modelling And Prediction in PhotoIonised Nebulae and Gasdynamical Shocks (MAPPINGS III) code. Among the options offered by ITERA, in \citet{Bonato2015} we chose, for starbursts, the \citet{Dopita06} models and, for AGNs, the dust free isochoric narrow line region (NLR) models for type\,1's and the dusty radiation-pressure dominated NLR models for type\,2's \citep{Groves04}. The chosen models were those which provided the best overall fit (minimum $\chi^2$) to the observed line ratios of local starbursts in the \citet{Bernard-Salas2009} catalogue and of AGNs in the sample built by \citet{Bonato2014b} combining sources from the \citet{Sturm02}, \citet{Tomm08,Tommasin2010}, and \citet{Veilleux09} catalogues. In \citet{Bonato2015}, our theoretical calibrations were compared with observational data, showing a very good agreement and so proving the goodness of our procedure.

To this initial sample we added the PAH bands at 3.3 and 12.7$\mu$m and the \textit{Pa}\,$\alpha$ (HI\,1.88$\mu$m) line.

\begin{figure}
\centering
  \includegraphics[trim=3.3cm 0.5cm 1.2cm 0.5cm,clip=true,width=0.47\textwidth]{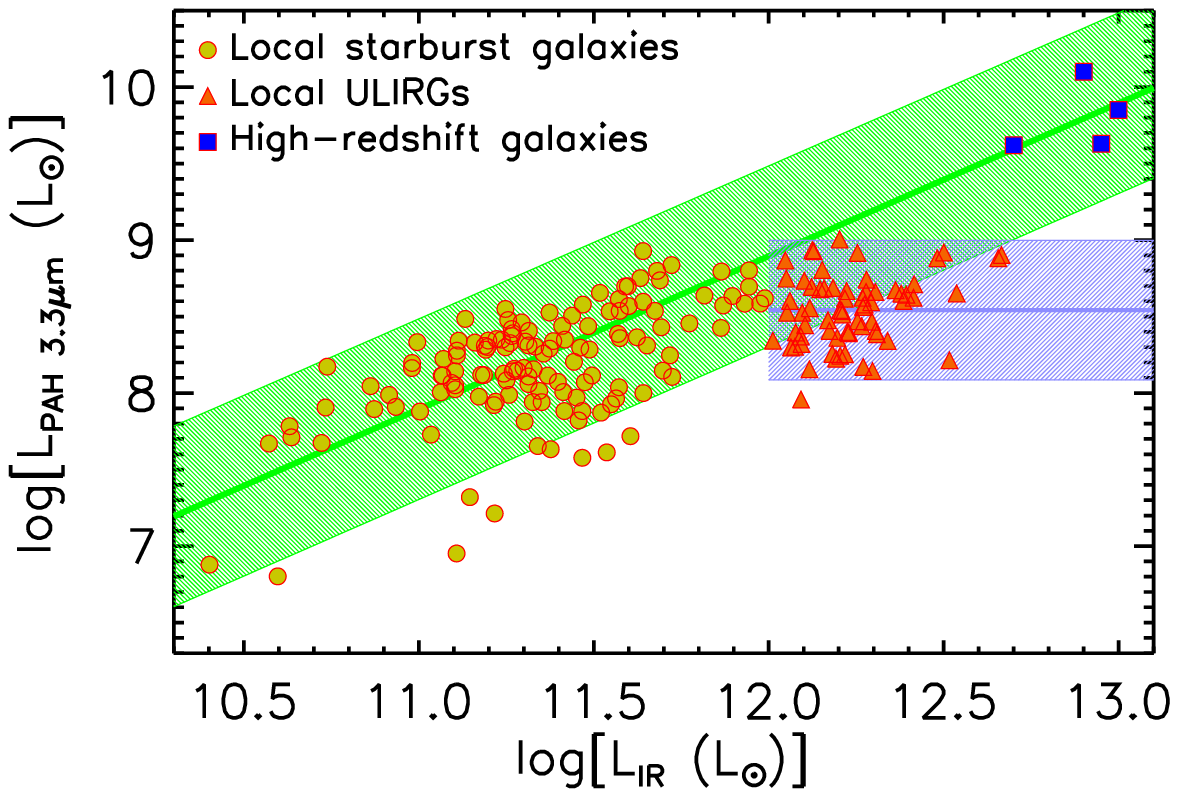} 
  \includegraphics[trim=2.8cm 0.5cm 1.6cm 0.5cm,clip=true,width=0.47\textwidth]{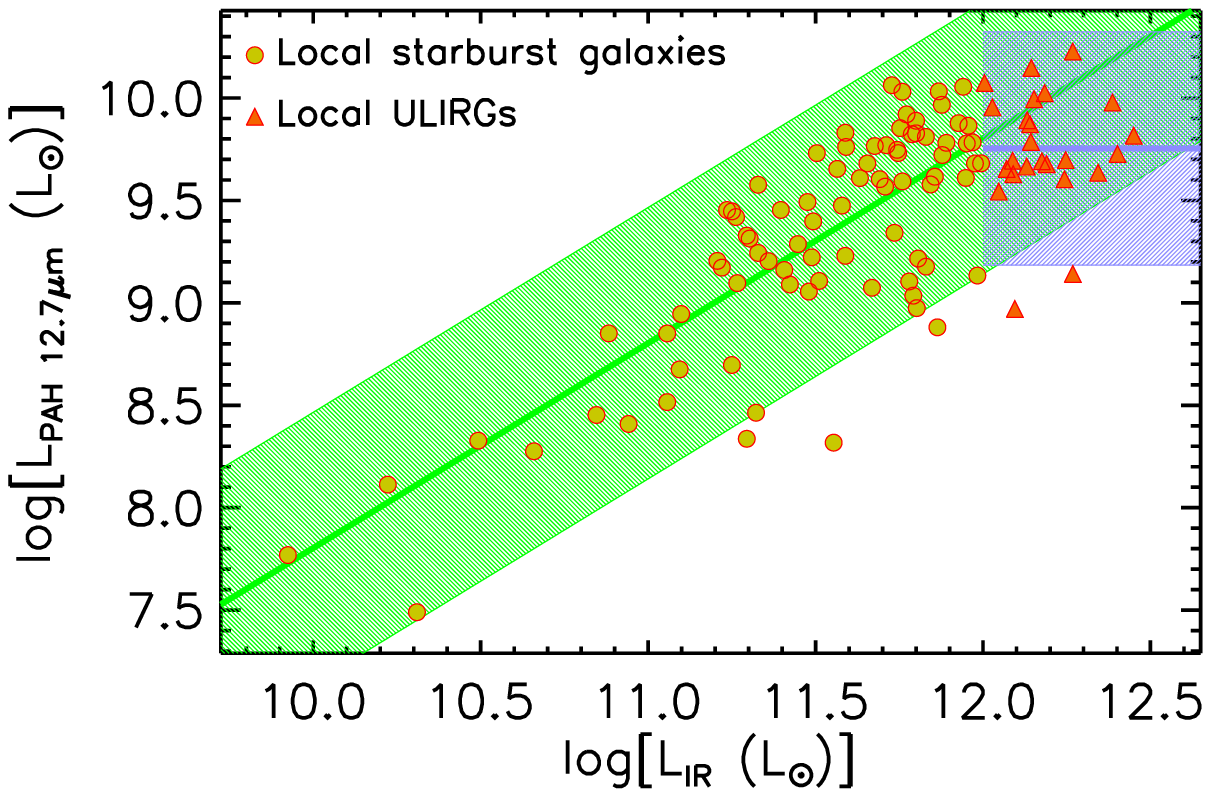}
\caption{Luminosity of the PAH\,3.3\,$\mu$m (\textit{left})/PAH\,12.7\,$\mu$m (\textit{right}) line versus continuum IR luminosity. The green bands show the $2\,\sigma$ range around the mean linear relation $\log(L_{\ell})=\log(L_{\rm IR}) + c$ for local star forming galaxies with $L_{\rm IR}<10^{12}\,L_{\sun}$ (circles) and high-redshift galaxies (squares; only for PAH\,3.3\,$\mu$m line, data of high-z sources were available); the derived value of $c\equiv\langle \log(L_{\ell}/L_{\rm IR})\rangle$ is $-3.11\pm0.30$ (for PAH\,3.3\,$\mu$m) and $-2.20\pm0.33$ (for PAH\,12.7\,$\mu$m). The azure bands show the $2\,\sigma$ spread around the mean line luminosity for the sample of local ULIRGs (triangles) whose line luminosities appear to be uncorrelated with $L_{\rm IR}$; the mean line luminosity $\langle \log L_{\ell}\rangle$ of these objects is $8.54\pm0.23$ (for PAH\,3.3\,$\mu$m) and $9.75\pm0.28$ (for PAH\,12.7\,$\mu$m).}
  \label{fig:PAH_cal}
\end{figure}

For the PAH 3.3$\mu$m line, we collected data (on line and continuum luminosity) of star-forming galaxies from: \citet{Rodriguez2003}, \citet{Imanishi2008,Imanishi2010}, \citet{Sajina2009}, \citet{Lee2012}, \citet{Kim2012} and \citet{Yamada2013}. We excluded objects for which there is evidence for a substantial AGN contribution to the total IR luminosity. The left panel of Fig.\,\ref{fig:PAH_cal} shows the 3.3$\mu$m PAH line luminosities vs. $L_{\rm IR}$ for both local and high-$z$ ($z\sim2$) sources. 

Excluding local ULIRGS ($L_{\rm IR}> 10^{12}\,L_\odot$, orange triangles), that are known to be under-luminous in PAH lines (e.g. \citealt{Bonato2014a} and references therein), sources show a highly significant linear correlation\footnote{In \citet{Bonato2014a}, considering star-forming (stellar/HII and photodissociation region) lines and purely star-forming (non ULIRG) galaxies, we exploited the relationship between line and continuum luminosity, with the support of extensive simulations (taking into account dust obscuration) and through the comparison with observational data. We showed that a direct proportionality between line and continuum luminosity (i.e. $\log(L_{\ell})=\log(L_{\rm IR}) + c$) is preferable to a linear proportionality with a free slope (i.e. $\log(L_{\ell})=a\times\log(L_{\rm IR}) + b$).}, with mean $\langle \log(L_\ell/L_{\rm IR} \rangle = -3.11\pm 0.30$ where the ``error'', $\sigma = 0.30$, is the dispersion around the mean.  The correlation coefficient is $\simeq 0.8$, corresponding to a probability of no correlation of $\simeq 10^{-127}$.

More than 50\% of the local ULIRGs lie on the right of the green band representing  the $\pm 2\,\sigma$  interval around the mean $\log(L_{\rm IR})$-$\log(L_{\rm PAH})$ relation holding for the other sources. Since local ULIRGs do not show any significant correlation between $\log(L_{\rm IR})$ and $\log(L_{\rm PAH})$, perhaps due to their limited luminosity range,  we adopt a Gaussian distribution of the logarithm of the line luminosity, $\log(L_\ell)$, 
around a mean value, $\langle \log(L_\ell) \rangle = 8.54$ with a dispersion of 0.23. Actually assigning PAH luminosities to local ULIRGs using either the correlation or the Gaussian distribution does not significantly affect our results: in both cases all their PAH lines are detected by pointed observations with  the minimum integration time considered here (see Subsect. 5.2), and they give a negligible contribution to the counts of serendipitously detected sources, as expected given their very low space density. We checked that this conclusion holds independently of the transition redshift between ``local'' and ``high-$z$'' ULIRGs by considering two extreme cases with transition redshifts of $z=0.2$ and $z=1.5$.

For the PAH\,12.7$\mu m$, we collected data of local star-forming galaxies from \citet{Shipley2013} and \citet{dasyra09}, shown on the right panel of Fig.\,\ref{fig:PAH_cal}. Also in this case, we excluded objects for which there is evidence for a substantial AGN contribution to the total IR luminosity. As done for the other PAH lines, we have first investigated the correlation between line and IR luminosities excluding local ULIRGs. Again the data are consistent with a linear relation; we find $\langle \log(L_\ell/L_{\rm IR} \rangle = -2.20\pm 0.33$. Although in this case local ULIRGs lie within the green band, for homogeneity with the treatment of the other PAH lines  we adopt a Gaussian distribution with $\langle \log(L_\ell) \rangle = 9.75$  and dispersion of 0.28.



Apart from local ULIRGs, there are indications in the literature of a PAH emission deficit in low-metallicity star-forming dwarf galaxies as well, compared to higher metallicity and/or higher luminosity star-forming galaxies (see e.g. \citealt{Engelbracht2005,Engelbracht2008}, \citealt{Wu2006}, \citealt{OHalloran2006}, \citealt{Madden2006}, \citealt{Smith2007}, \citealt{Rosenberg2008}, \citealt{Galametz2009}). Most of these studies suggest a decrement by a factor $\sim$10 in the L$_{\rm PAH}$/SFR ratio, probably due to the fact that low-metallicity dwarf galaxies have very small grains and/or more hot dust. In particular, the lower PAH emission could be caused by the ability of the interstellar radiation field to penetrate lower dust column densities (lower dust-to-gas ratios are expected in low-metallicity conditions), dissociating or destroying the PAHs (\citealt{Galliano2003}).
We tested the consistency of our results considering the case of a deficit equal to 10 in PAH luminosity in all our simulated galaxies having $L_{\rm IR}\leq10^{9}\,L_{\sun}$\footnote{In \citealt{Bonato2014a}, our Fig.5, we showed, for several metal lines, the decrement of the line-continuum luminosity ratio, at $L_{\rm IR}\leq10^{9}\,L_{\sun}$, probably due to metallicity. Similar results can be obtained from the mass-metallicity relation (see e.g. \citealt{Tremonti2004}; \citealt{Zahid2013}).} (at all z; see Subsect.\,\ref{subsect:MRS_ss}). 

For the \textit{Pa}\,$\alpha$ line, because of the scarcity of observational data useful for an empirical calibration, we adopted the same procedure for theoretical calibrations described above. For star-forming galaxies, we obtained $\langle \log(L_\ell/L_{\rm IR} \rangle = -3.32\pm 0.30$. This value is consistent, for example, with the relation adopted by \citealt{Rieke09}, i.e. $\log({L_{\textit{Pa}\,\alpha}/L_{\rm IR}})\sim-3.14$, who used the \citet{Kennicutt1998} $SFR-L_{\rm IR}$ relation and a \textit{Pa}\,$\alpha$/\textit{H}\,$\alpha$ ratio of 0.128 (\citealt{Hummer87}). We derived, as coefficients of the best-fit linear relations between line and AGN bolometric luminosities, $\log({L_{\ell}})=a\cdot\log({L_{\rm bol}})+b$, and $1\sigma$ dispersions associated to the relation: $a=0.85$, $b=-2.21$, $disp=0.34$.

We verified that our calibrations between line and continuum luminosity are consistent with the SED library adopted in \citet{Kurinsky2016}. In particular, for the 6 PAH features included in our line sample, we compared the $c=\langle \log(L_{\ell}/L_{\rm IR})\rangle$ of our calibrations with the mean values derived from the star-forming galaxy templates adopted in \citet[taken from \citealt{Rieke09} and \citealt{Kirkpatrick2015}]{Kurinsky2016}. From these SEDs, using the IDL tool PAHFIT (\citealt{Smith2007}), we obtained the following mean quantities (compared in parenthesis with the 1-$\sigma$ lower and upper limits of our calibrations): c$_{PAH\,3.3\mu m}=-3.12\pm0.54$ ([$-3.41$,$-2.81$]); c$_{PAH\,6.2\mu m}=-2.10\pm0.12$ ([$-2.56$,$-1.84$]); c$_{PAH\,7.7\mu m}=-1.43\pm0.12$ ([$-2.00$,$-1.28$]); c$_{PAH\,8.6\mu m}=-2.20\pm0.14$ ([$-2.52$,$-1.80$]); c$_{PAH\,11.3\mu m}=-2.07\pm0.20$ ([$-2.65$,$-1.93$]); c$_{PAH\,12.7\mu m}=-2.31\pm0.11$ ([$-2.53$,$-1.87$]).

Apart from the PAH lines, all the other line calibrations are based on local observations. In this paper, we assume that these relations are not subject to redshift evolution. Indeed, MIRI will be the first instrument which will be able to test this assumption explicitly.

The line luminosity functions of all the lines of our sample have been computed starting from the redshift-dependent IR luminosity functions given by the evolutionary models. To properly take into account the dispersion in the relationships between line and continuum luminosities, we used the Monte Carlo approach described in \citet{Bonato2014a}.

\section{MIRI MRS predictions}\label{sect:MRS_surveys}

\subsection{Pointed observations}\label{subsect:MRS_po}

\begin{figure}
\centering
    \includegraphics[trim=2.4cm 0.6cm 3.3cm 0.6cm,clip=true,width=0.37\textwidth, angle=0]{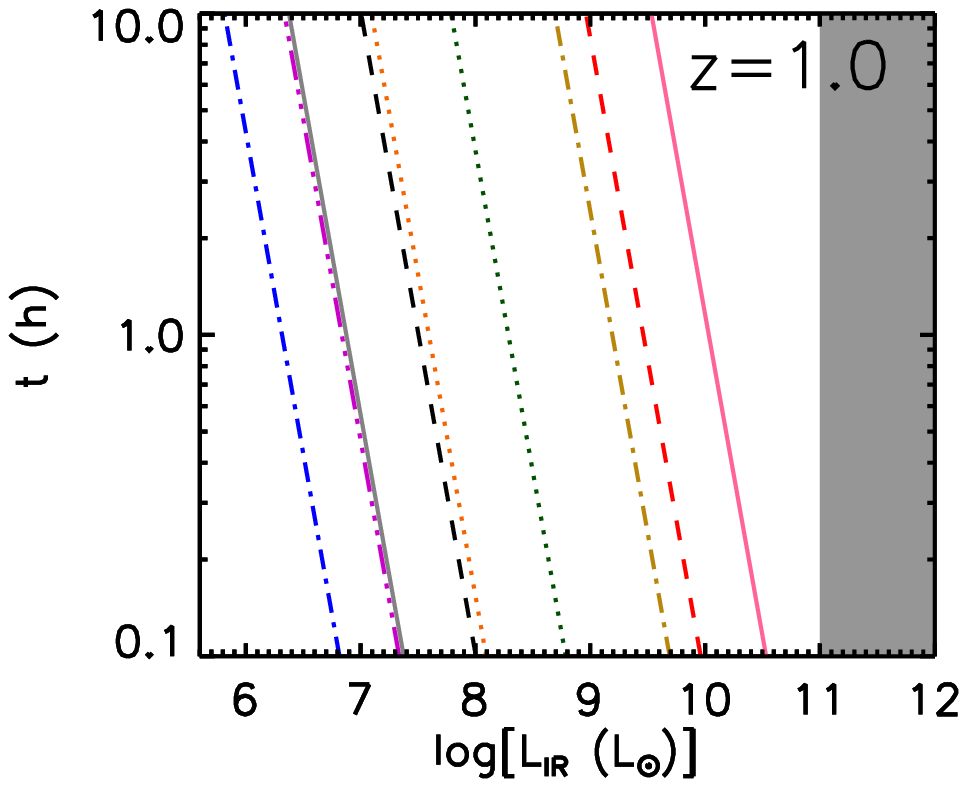}
    \includegraphics[trim=2.4cm 0.6cm 0.0cm 0.6cm,clip=true,width=0.48\textwidth, angle=0]{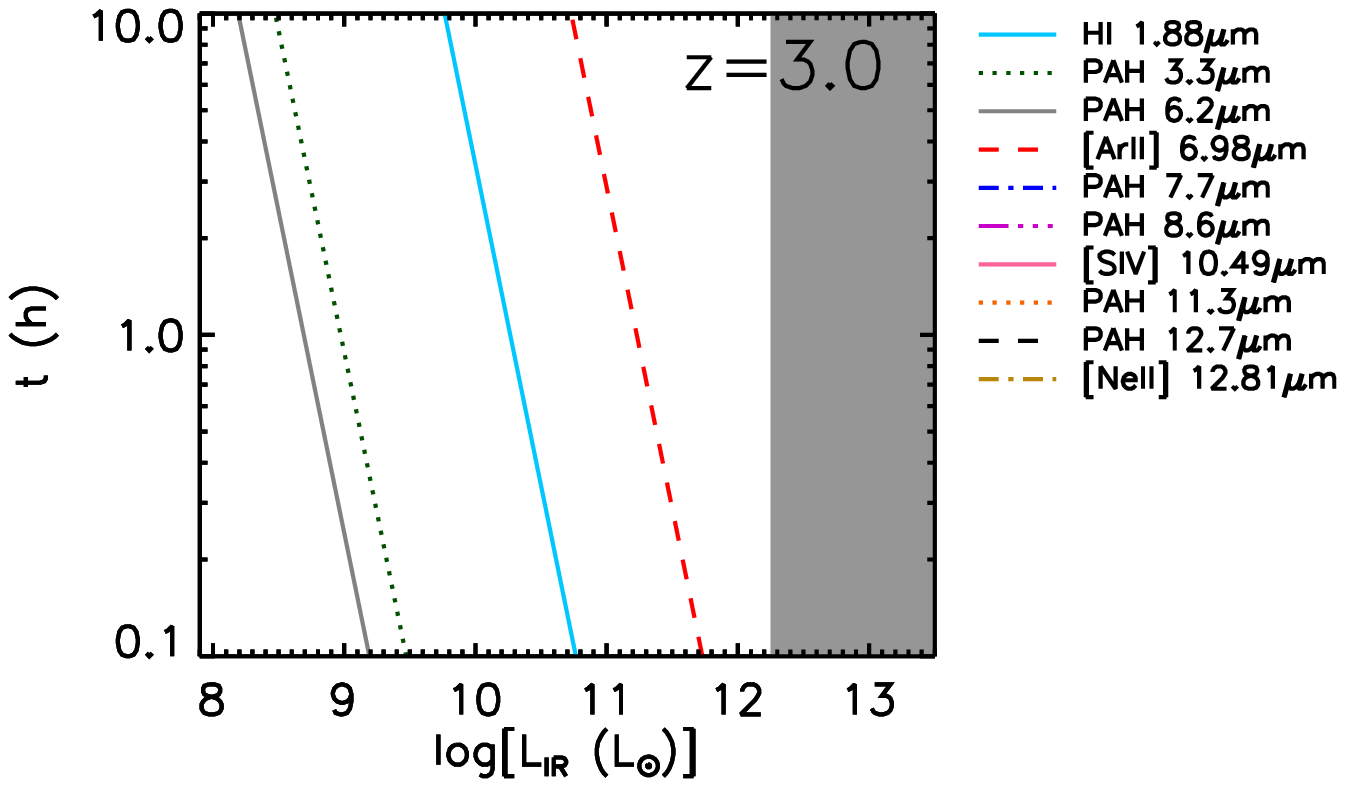}
		
  \caption{MIRI MRS exposure time vs. minimum IR luminosity (of star-forming galaxies) for a given line to be detected at 5$\sigma$. We show both the predictions for sources at $z=1$ (\textit{left}) and $z=3$ (\textit{right}). In both cases, the shaded regions indicate the regime of \textit{Herschel}/PACS and SPIRE-detected sources (based on \citealt{Gruppioni2013}). All the lines to the left of the shaded regions will be detected within the minimum time shown (i.e. 0.1\,h) in the \textit{Herschel} sources. Only the key star-forming lines are shown for clarity. Almost all the omitted spectral lines (H$_{2}$\,5.51, H$_{2}$\,6.91, [ArIII]\,8.99,  H$_{2}$\,9.66, H$_{2}$\,12.28,  HI\,12.37, [ClII]\,14.38 and Sil 9.7$\,\mu$m) are fainter than the lines shown in the figure.}
  \label{fig:strategy_MRS_SF}
\end{figure}

\begin{figure}
\centering
    \includegraphics[trim=2.4cm 0.6cm 3.5cm 0.8cm,clip=true,width=0.37\textwidth, angle=0]{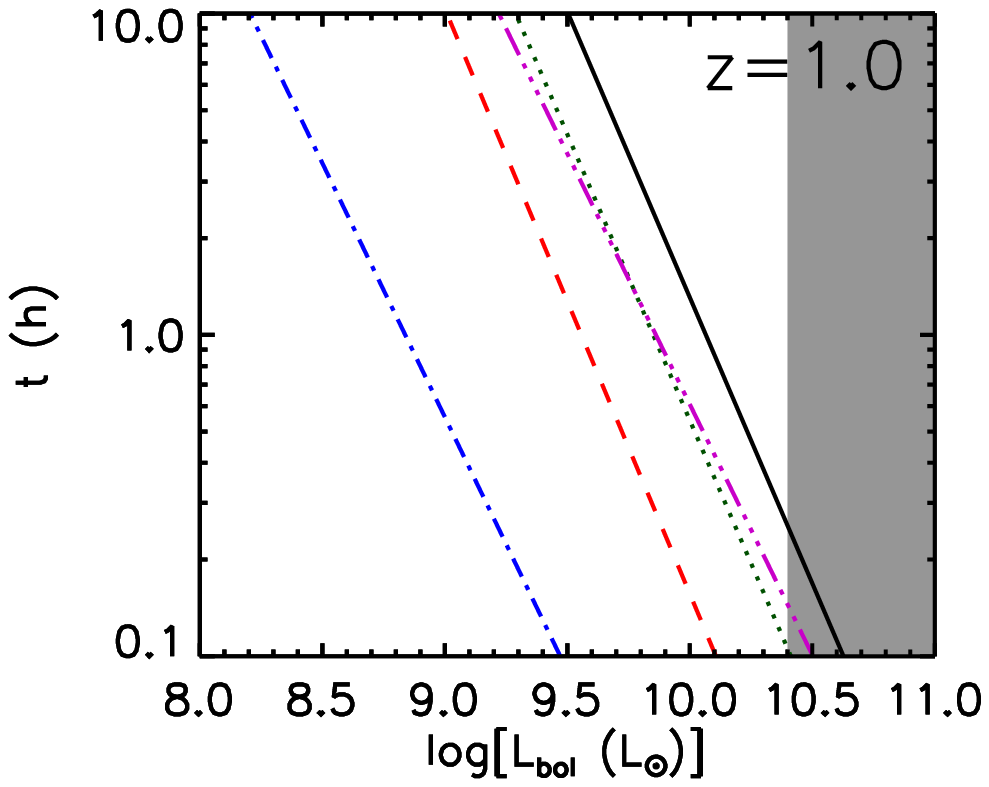}
    \includegraphics[trim=0.5cm 0.6cm 0.0cm 0.8cm,clip=true,width=0.56\textwidth, angle=0]{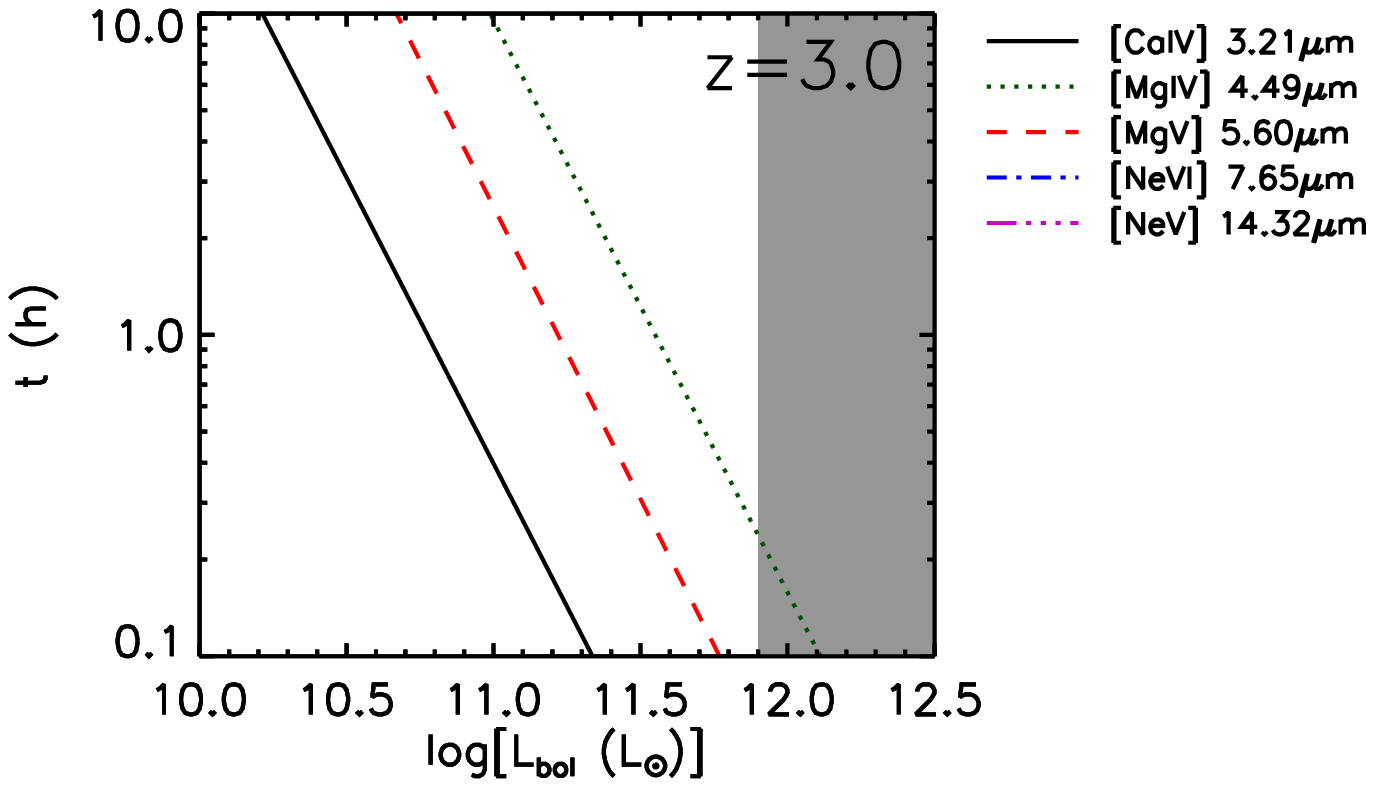}
    
  \caption{MIRI MRS exposure time vs. minimum bolometric luminosity (of AGN) for a given typical AGN line to be detected at 5$\sigma$. We show both the predictions for sources at $z=1$ (\textit{left}) and $z=3$ (\textit{right}). In both cases, the shaded regions indicate the regime of \textit{Herschel}/PACS and SPIRE-detected AGN sources (based on \citealt{Delvecchio2014}). All the lines to the left of the shaded regions will be detected within the minimum time shown (i.e. 0.1\,h) in the \textit{Herschel} sources. Only the key AGN lines are shown for clarity. The omitted AGN lines ([MgVIII]\,3.03, [SiIX]\,3.92, [CaV]\,4.20, [ArVI]\,4.52, [SiVII]\,6.50, [ArV]\,7.90, [CaV]\,11.48, [ArV]\,13.09 and [MgV]\,13.50$\,\mu$m) are fainter than the lines shown in the figure.}    
  \label{fig:strategy_MRS_AGN}
\end{figure}

It is likely that some of the earliest pointed observations with the MRS will focus on the IR-luminous dust obscured populations such as the \textit{Herschel}-selected galaxies (e.g. \citealt{Elbaz2011}, \citealt{Oliver2012}, \citealt{Gruppioni2013}, \citealt{Eales2010}, \citealt{Valiante2016}, \citealt{Bourne2016}). Figs.~\ref{fig:strategy_MRS_SF} and ~\ref{fig:strategy_MRS_AGN} show the MRS exposure times\footnote{The minimum integration time available for the MRS is $\sim$30\,s (SLOWMode readout time, \citealt{Ressler2015}). In this paper we consider a ``minimum'' exposure time of 0.1\,h, because observations having a lower integration time would be very disadvantageous in term of overheads (see Tab.\,6 in \citealt{Bouchet2015}; see also \citealt{Gordon2015}) and so we expect that it is unlikely they will be planned.} needed to achieve a $5\,\sigma$ detection of SF and AGN lines at $z=1$ and $z=3$ as a function of the IR luminosity (of the SF component) and of the bolometric luminosity (of the AGN component), respectively. The shaded areas correspond to the IR luminosities represented in the SF luminosity functions determined by \citet{Gruppioni2013} on the basis of \textit{Herschel}/PACS and SPIRE surveys, and to the bolometric luminosities represented in the AGN bolometric luminosity functions derived by \citet{Delvecchio2014} on the basis of the same surveys, respectively.
 
The exposure times for the PAH and the silicate bands have been computed degrading the resolution in the four MRS channels by a factor of 58, 46, 45 and 36, respectively, to obtain $R\simeq 60$. The suitability of such resolution for PAH detections has been already demonstrated from the analysis of {\sl Spitzer} low-resolution spectra (see e.g. \citealt{Fleming2010}, \citealt{Fiolet2010}, \citealt{ODowd2011}). For the fine-structure lines we considered the full resolution spectra.  In other words, we envisage that the full resolution spectra are analyzed first to look for fine-structure lines and then are degraded to increase the number of PAH and silicate band detections.
 
In Fig.~\ref{fig:strategy_MRS_SF} we see that in a few minutes all the $z=1$ \textit{Herschel} galaxies will be detected in several (at least 9) star-forming lines while at $z=3$ they will be detected in 4 lines\footnote{Note that not only the \textit{Herschel} star-forming galaxies will be detected in these lines, but also the \textit{Herschel} AGNs. As pointed out by \citet{Gruppioni2013}, the SF component dominates the total IR luminosity of the \textit{Herschel} sources, even in the majority of AGNs. At e.g. z$\sim$3, if we consider the faintest \textit{Herschel} AGNs (having $L_{\rm IR}\sim2\times10^{12} L_{\sun}$) with a minimum SF component of 50\%, its SF-only $L_{\rm IR}$ is higher than the minimum $L_{\rm IR}$ achieved by the MRS in the PAH\,3.3/6.2$\mu$m, \textit{Pa}\,$\alpha$ and [ArII]\,6.98\,$\mu$m lines.},  thus providing spectroscopic redshift measurements for all of them. Fig.~\ref{fig:strategy_MRS_AGN} shows that in the same time the \textit{Herschel} sources will be detected in at least two pure AGN lines, allowing for the strength of the AGN to be assessed. The implications of these detections are discussed further in Section\,\ref{sec:discussion}.

\subsection{Serendipitous detections}\label{subsect:MRS_ss}

\begin{figure}
\centering
\makebox[\textwidth][c]{
\includegraphics[trim=3.1cm 0.5cm 1.1cm
    0.7cm,clip=true,width=0.43\textwidth, angle=0]{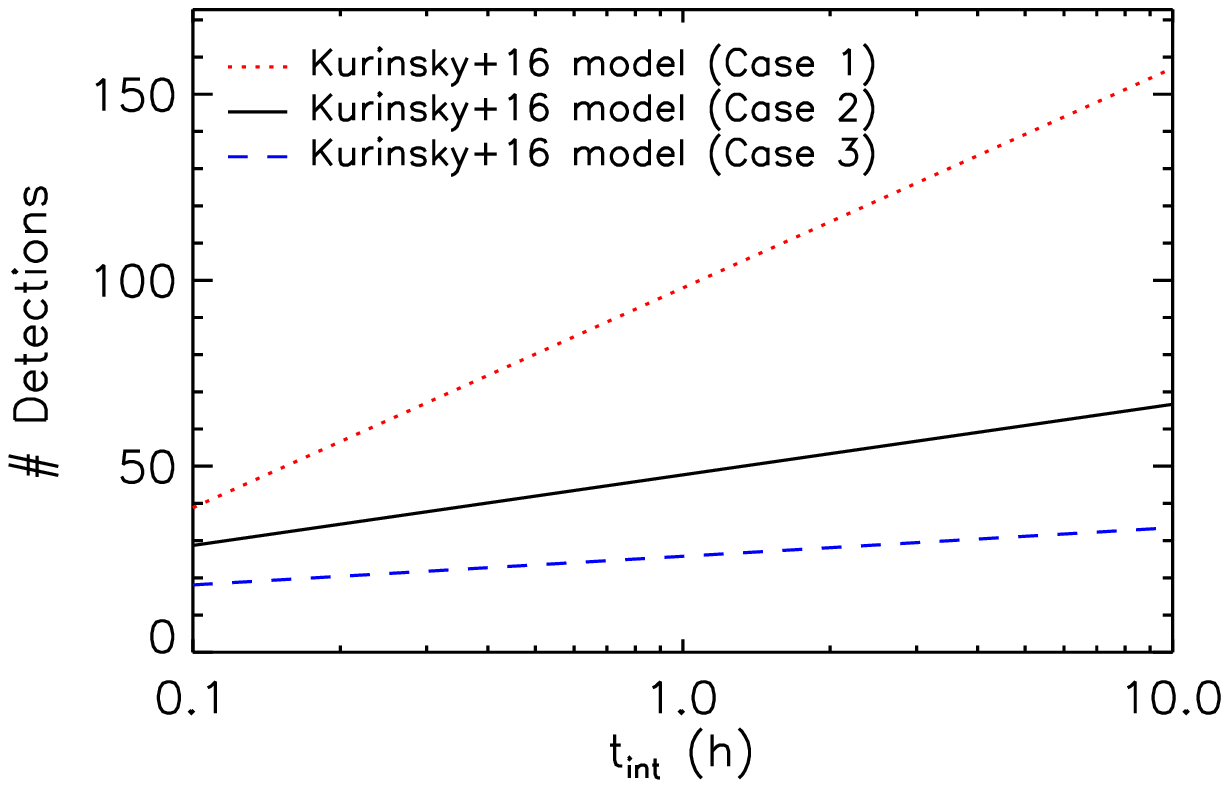}
\includegraphics[trim=3.1cm 0.5cm 1.1cm
    0.7cm,clip=true,width=0.43\textwidth, angle=0]{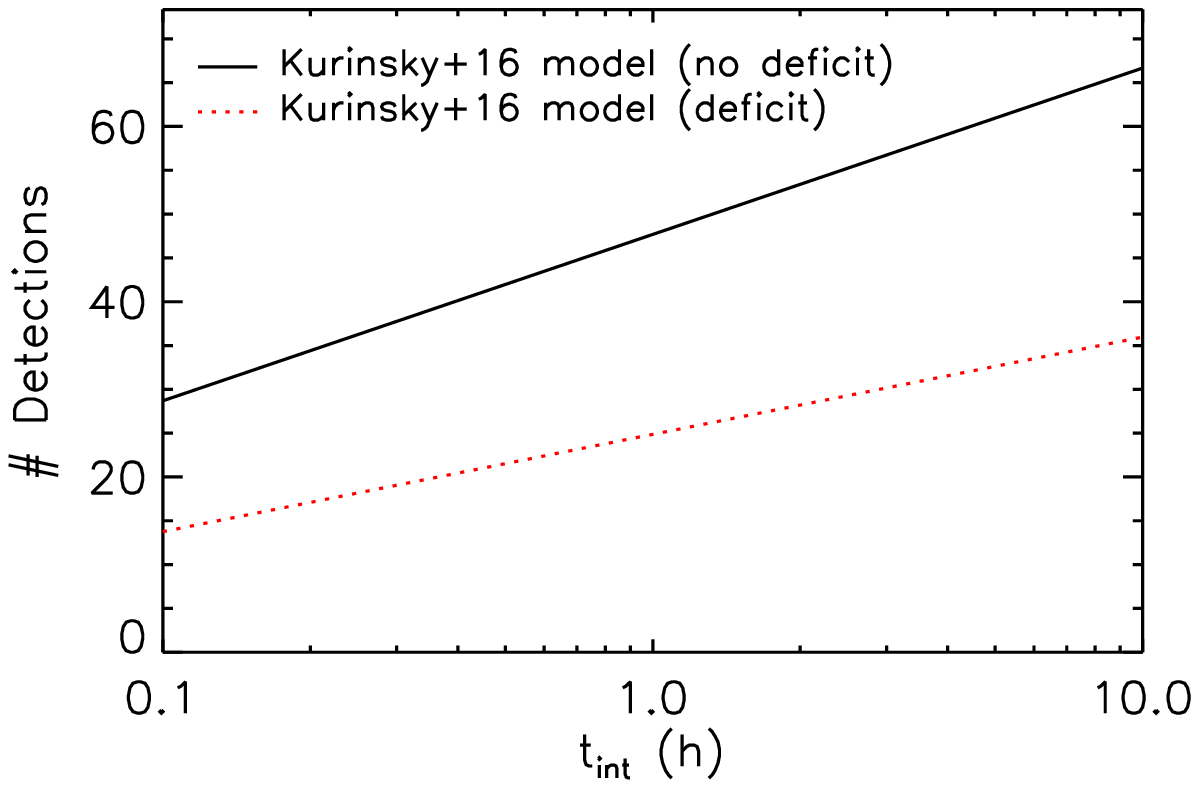}
}
  \caption{Number of serendipitous 5$\sigma$ line detections per FoV, as a function of the integration time. \textit{Left:} 
  for the \citet{Kurinsky2016} model, comparison between results obtained using the 3 different assumptions for the low-$L$ end of the IR LF. \textit{Right:} for the reference model (\citealt{Kurinsky2016} model with case 2 low-$L$ end assumption), comparison between detections obtained using the reference correlations between L$_{\rm IR}$ and L$_{\rm PAH}$ and those obtained using the assumption of a factor 10 deficit in L$_{\rm PAH}$ for all the galaxies having $L_{\rm IR}<10^{9}\,L_{\sun}$.}
  \label{fig:MRS_detections}
\end{figure}

Here we consider the question -- for a given exposure time, how many additional (serendipitous) sources should be detected in the same MRS FoV? Fig.~\ref{fig:MRS_detections} shows the number of 5$\sigma$ line detections, in a single exposure, as a function of the integration time. 
The left panel of Fig.~\ref{fig:MRS_detections} 
shows the effect of the different assumptions about the faint end of the LFs. On the right, we show how an extreme assumption about the PAH emission of low luminosity galaxies - simulated galaxies with $L_{\rm IR}<10^{9}\,L_{\sun}$ are under-luminous by a factor of 10 in the PAH lines - affects the results.
Globally, the number of detections per FoV varies from a few tens (with 0.1\,h integration) to tens/a few hundreds (with 10\,h). Testing these results by the use of the \citet{Cai13} model we found differences within only about a factor of 2 in the numbers obtained. Even for the shallowest observations, combining a few pointed observations of \textit{Herschel}-selected sources would yield statistically significant samples of serendipitous detections. These serendipitous studies will allow us to discriminate between the different scenarios - i.e. give us insight into the number density and properties of low-$L$ galaxies.

Most ($>$83\% in all the cases) of the detected lines are PAHs, as direct consequence of their brightness, of the much greater abundance of star-forming galaxies (compared to AGN) and of the degraded spectral resolution. The fraction of AGN detections (i.e. detections of pure AGN lines) is always $<$6\%.
The implications of these detections are discussed further in Section\,\ref{sec:discussion}.

\section{Discussion} \label{sec:discussion}

\subsection{Implications for the study of the galaxy-AGN co-evolution}

The study of galaxy-AGN (co-)evolution ultimately requires two different yet complementary approaches. First, we need to understand the detailed physical conditions in galaxies while they are actively star-forming and growing their central black holes. Next we need better constraints on the cosmic history of SF and black hole accretion. 

\subsubsection{The role of AGN in Herschel-selected galaxies \label{sec:agndiag}}
Figs.~\ref{fig:strategy_MRS_SF}$-$\ref{fig:strategy_MRS_AGN} show that, in $\sim$0.1\,h, all \textit{Herschel}-selected sources will be detected by MRS pointed observations in multiple lines, thus providing spectroscopic redshift measurements for all of them. The relative roles of AGN and SF in powering all \textit{Herschel}-selected galaxies can be assessed across this full redshift range by considering the PAH equivalent widths \citep[e.g.][]{Armus2007,Sajina2007,dasyra09}. For wide redshift ranges, these can be cross-calibrated using ratios of AGN to SF powered fine-structure lines, that provide more accurate estimates than PAH EWs. Among the best AGN diagnostic lines for the MRS is [NeVI]\,7.65$\mu$m, a strong AGN line (see \citealt{Spinoglio1992}), detectable over a broad redshift
range from z$=$0 to z$\sim$2.8. Within this redshift range, detecting the [NeVI]\,7.65$\mu$m line in \textit{Herschel}-selected sources requires only $\sim$0.1\,h. A commonly-used SF diagnostic line is [NeII]\,12.81$\mu$m which can be detected in any {\sl Herschel} source by the MRS in the range 0$<$z$\lesssim$\,1.2 in $\sim$0.1\,h. Beyond this redshift, up to z$\sim$3.1, and with the same integration time, we can use the [ArII]\,6.98\,$\mu$m SF line. 

In $\lesssim$1\,h, the [NeIII]15.55$\mu$m /[NeII]\,12.81$\mu$m vs. [SIV]\,10.49$\mu$m/[NeII]\,12.81$\mu$m diagnostic plot can be applied to all \textit{Herschel}-galaxies with $z\lesssim$0.9. This plot can be used to model age, metallicity and ionization parameter of star-forming galaxies (see e.g. \citealt{Inami2013}). These line ratios can also be used (as in \citealt{Inami2013}) to constrain photoionization models, e.g. using the MAPPINGS III photoionization code (\citealt{Binette85}, \citealt{Sutherland93}, \citealt{Groves04}) in combination with the Starburst99 stellar population synthesis code (\citealt{Leitherer99}, \citealt{Vazquez2005}). 

\subsubsection{The cosmic star-formation and black hole accretion rate histories and the low-L end of the IR luminosity function}

\begin{figure} 
\centering
    \includegraphics*[trim=1.5cm 0.6cm 0.5cm 1.0cm,clip=true,width=0.7\textwidth, angle=0]{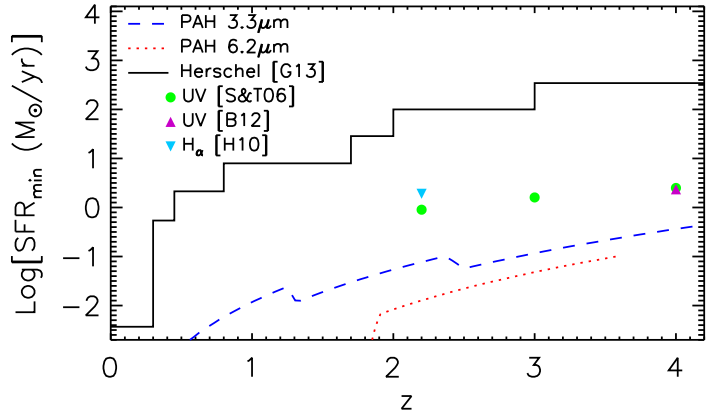}
  \caption{As a function of the redshift, comparison between the minimum SFR achieved by a MRS serendipitous survey with 0.1\,h integration (through spectroscopic detections of PAH 3.3 and 6.2$\mu$m lines) and the minimum SFR reached by other IR (SFR corresponding to the minimum luminosities represented in the IR luminosity functions determined by \citet{Gruppioni2013} [G13] on the basis of \textit{Herschel}/PACS and SPIRE surveys), H$_{\alpha}$ (\citealt{Hayes10} [H10]) and UV (\citealt{Sawicki06} [S\&T06], \citealt{Bouwens12} [B12]) measurements. The minimum SFR achieved by the MRS in the two PAH 3.3 and 6.2$\mu$m lines was calculated considering the detection limit at 0.1\,h integration, our calibrations between PAH and IR luminosity and the $L_{\rm IR}$/SFR relation derived by \citet{Clemens2013}.}
  \label{fig:SFR}
\end{figure}

Serendipitously detected galaxies could be used to trace the SFR history.
Figure~\ref{fig:SFR} shows the minimum SFR (calculated using our line/IR luminosity relations and the $L_{\rm IR}$/SFR relation by \citealt{Clemens2013}) of the sources detectable by the MRS (for 0.1\,h, the ``minimum'' exposure time needed for the detection of \textit{Herschel} sources). Also shown, for comparison, are the minimum SFR detected in existing {\sl Herschel} PACS+SPIRE surveys \citep{Gruppioni2013}. The improvement over \textit{Herschel} is impressive, about 3 orders of magnitude. Comparison with the lowest SFR achieved by the deepest H$_{\alpha}$ (\citealt{Hayes10}) and UV (\citealt{Sawicki06}, \citealt{Bouwens12}) surveys at high-$z$ also shows a significant improvement achievable through the MRS serendipitous detections. Using such survey data we will sample SFRs well below those of the most efficient star formers at the peak of the cosmic star formation activity, at $z$ between 2 and 3, estimated to be $\simeq 100\,M_{\sun}$/yr \citep{Genzel06,Cai13}. 

\begin{figure} 
\centering
    \includegraphics*[trim=2.5cm 2.5cm 3.0cm 3.0cm,clip=true,width=0.99\textwidth, angle=0]{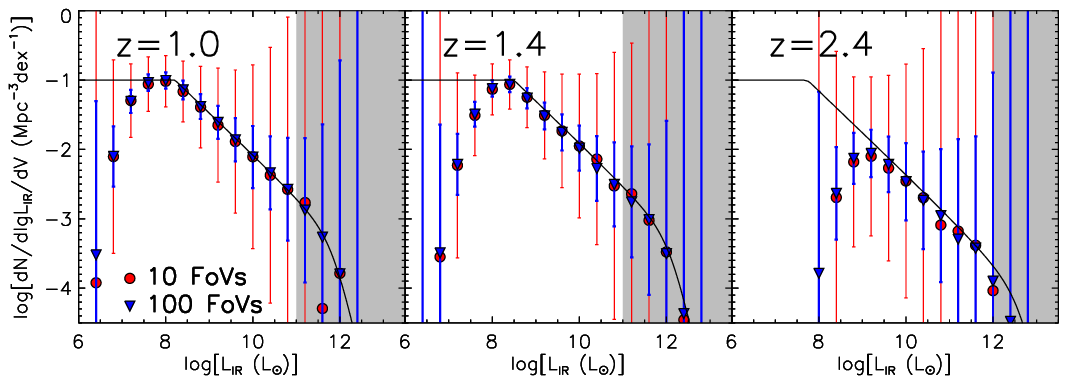}
  \caption{Adopting the reference model (``intermediate case'' for the \citealt{Kurinsky2016} model), at three different redshifts, comparison between the model IR LFs (of the star-forming population; black lines) and simulated reconstructions of them, using 10 (red circles) and 100 (blue triangles) FoVs of MRS serendipitous detections with 0.1\,h integration. The shaded regions indicate the regime of \textit{Herschel}/PACS and SPIRE-detected sources (based on \citealt{Gruppioni2013}).}
  \label{fig:rec_SF}
\end{figure}

For a good reconstruction of the IR LFs, and therefore of the SFR history, through serendipitous spectroscopic detections in 0.1\,h integration, we find that $\sim$100 FoVs will be necessary. For the reference model (``intermediate case'' of the \citealt{Kurinsky2016} model), Fig.~\ref{fig:rec_SF} shows IR LFs (of the star-forming population) reconstructed, at three different redshifts, using serendipitous sources collected from 10 and 100 FoVs and detected in at least 2 spectral lines.

The LFs in the different bins in redshift and IR luminosity (adopting bin sizes $\Delta z=0.2$ and $\Delta \log L=0.4$) were derived using the $1/V_{\rm max}$ method \citep{Schmidt1968}:
\begin{equation}
{dN(L_j,z_k)\over d\log L}=\frac{1}{\Delta \log L}\sum_{i=1}^{N_{j}} \frac{1}{V_{{\rm max},i}(z_k)}
\end{equation}
where $z_k$ is the bin center and the sum is over all the $N_{j}$ sources with luminosity in the range [$\log L_{j}-\Delta\log L/2$, $\log L_{j}+\Delta\log L/2$] within the redshift bin. The quantity $V_{{\rm max},i}$ is the comoving volume, within the solid angle of the survey, enclosed between the lower ($z_{\rm min}$) and the upper ($z_{\rm max}$) limit of the bin. The mean $dN(L_j,z_k)/d\log L$ obtained from the simulations was adopted as the maximum likelihood value. The dispersions around the mean $dN(L_j,z_k)/d\log L$ were taken as estimates of the Poisson errors, calculated as: $\sigma_{j,k} = \left[\sum_{i=1}^{N_{j}} \left(\frac{1}{V_{{\rm max},i}}(z_k)\right)^2\right]^{1/2}$.

Note that the contribution of very low-L sources (placed in the regions of our three different model LF low-L end) to the star formation rate density (SFRD) is minor (see \citealt{Madau14}). 
However these serendipitous surveys will be able to explore levels of SFR never achieved before. They will provide an impressive improvement in the reconstruction of the faint end of the PAH line LFs, and therefore of the IR LFs and of the SFR functions.

Considering the simulated reconstructions of the LFs described above and using the first 7 bins in redshift (until z$=$1.4, i.e. where the incompleteness\footnote{This incompleteness effect is strictly correlated to the dispersion in the L$_{line}$-L$_{IR}$ relationships, i.e. to the fact that, close to the spectroscopic detection limit, only a fraction of the galaxies with a certain L$_{IR}$ has a L$_{line}$ greater than such detection limit.} does not affect significantly the measurement, see Fig.~\ref{fig:rec_SF}), we obtain a value for the slope of $\sim$0.1$\pm$0.3 with 100 FoVs. With a smaller number of FoVs, the uncertainty on this measurement would be too high (e.g. we obtain a slope $\sim$0.2$\pm$1.9 with 10 FoVs). But about one hundred of FoVs will provide a tight estimation of the slope of the low-L end.

\begin{figure} 
\centering
    \includegraphics*[trim=1.3cm 0.6cm 0.5cm 1.0cm,clip=true,width=0.7\textwidth, angle=0]{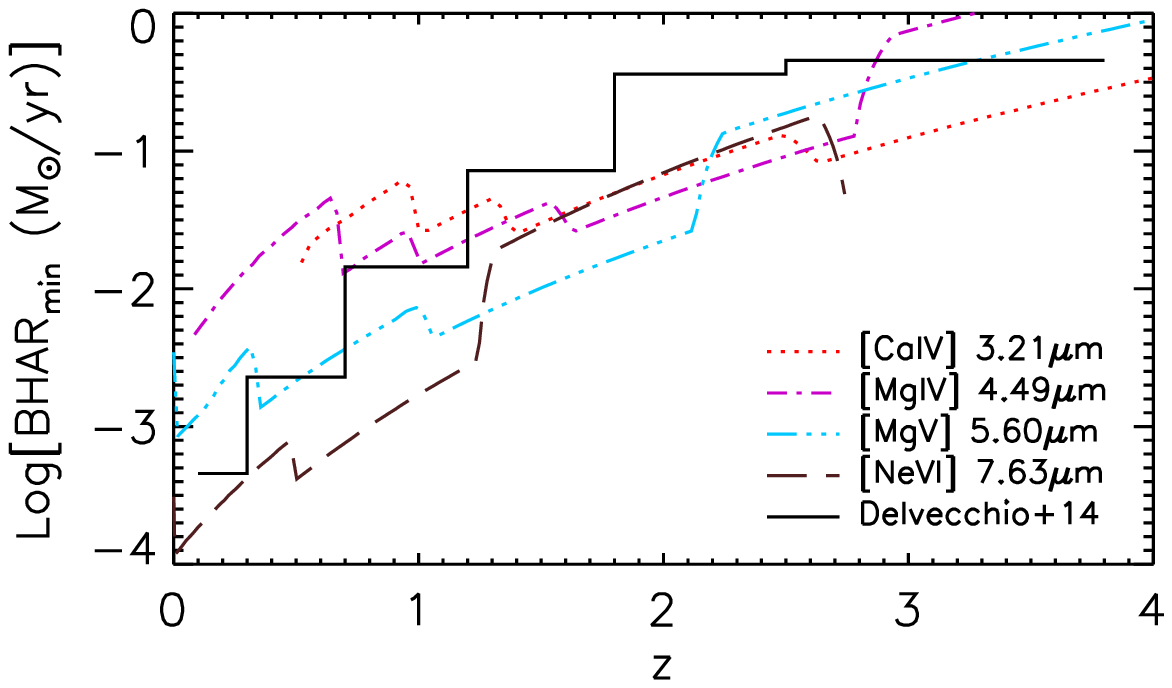}
  \caption{Comparison between the minimum BHAR achieved by a MRS serendipitous survey with 0.1\,h integration (through spectroscopic detections of several AGN lines) and the BHAR corresponding to the minimum bolometric luminosities represented in the AGN bolometric luminosity functions determined by \citet{Delvecchio2014} on the basis of \textit{Herschel}/PACS and SPIRE surveys, as a function of the redshift. The minimum BHAR achieved by the MRS in these four different AGN lines was calculated considering the detection limit at 0.1\,h integration, our calibrations between line and AGN bolometric luminosity, and the $L_{\rm bol}$/BHAR relation derived by \citealt{Chen2013}.}
  \label{fig:BHAR}
\end{figure}

In Fig.~\ref{fig:BHAR} we compare, as a function of redshift, the minimum black-hole accretion rates (BHARs; calculated using our line/bolometric luminosity relations and the $L_{\rm bol}$/BHAR relation by \citealt{Chen2013}) of the sources detectable by the MRS (again in 0.1\,h integration) with those associated to the minimum bolometric luminosities represented in the AGN bolometric luminosity functions determined by \citet{Delvecchio2014} on the basis of \textit{Herschel}/PACS and SPIRE surveys. We find an improvement over \textit{Herschel} of about 1 order of magnitude.

\begin{figure} 
\centering
    \includegraphics*[trim=2.5cm 2.5cm 3.0cm 3.0cm,clip=true,width=0.99\textwidth, angle=0]{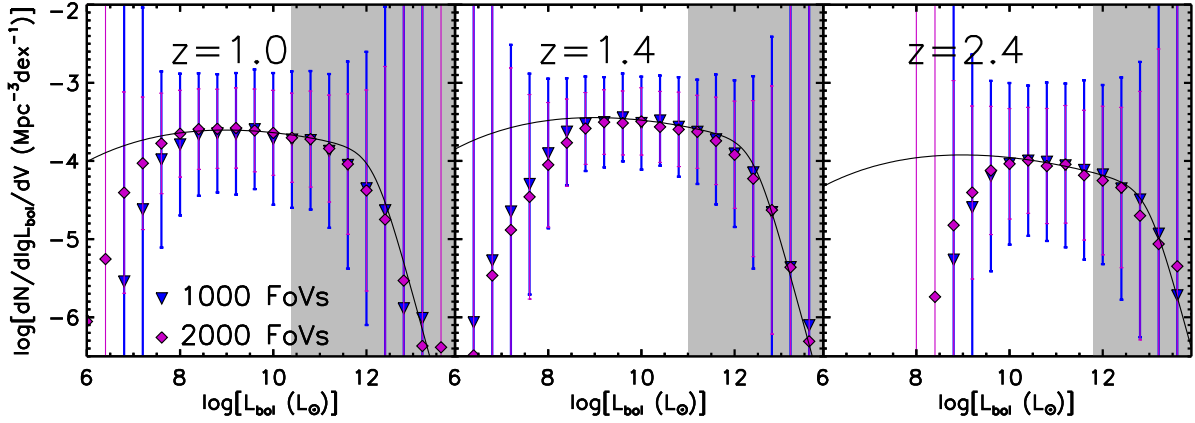}
  \caption{Adopting the reference model (``intermediate case'' for the \citealt{Kurinsky2016} model), at three different redshifts, comparison between the model AGN bolometric LFs (black lines) and simulated reconstructions of them, using 1000 (blue triangles) and 2000 (violet diamonds) FoVs of MRS serendipitous detections with 0.1\,h integration. The shaded regions indicate the regime of \textit{Herschel}/PACS and SPIRE-detected sources (based on \citealt{Delvecchio2014}).}
  \label{fig:rec_AGN}
\end{figure}

Despite this improvement in sensitivity, because of the relative rarity of AGNs, for a good reconstruction of the AGN bolometric LFs, and therefore of the BHAR functions, again with 0.1\,h integration, a large number of FoVs is necessary, i.e. $\gtrsim$1000, according to our simulations. Fig.~\ref{fig:rec_AGN} shows AGN bolometric LFs reconstructed, at three different redshifts, using serendipitous sources collected from 1000 and 2000 FoVs and detected in at least 2 spectral lines.

\begin{figure}
\centering
\makebox[\textwidth][c]{
\includegraphics[trim=2.8cm 0.3cm 1.3cm 0.2cm,clip=true,width=0.43\textwidth, angle=0]{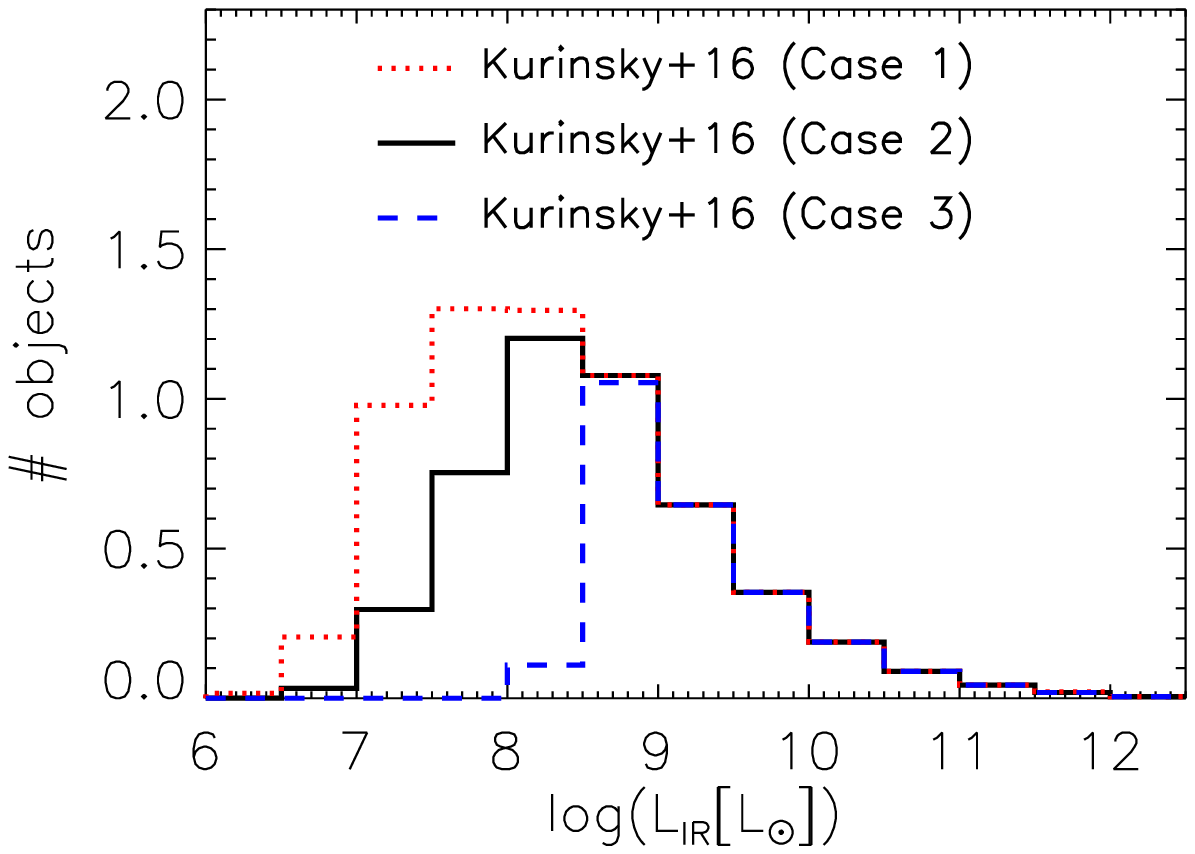}
\includegraphics[trim=2.8cm 0.3cm 1.3cm 0.2cm,clip=true,width=0.43\textwidth, angle=0]{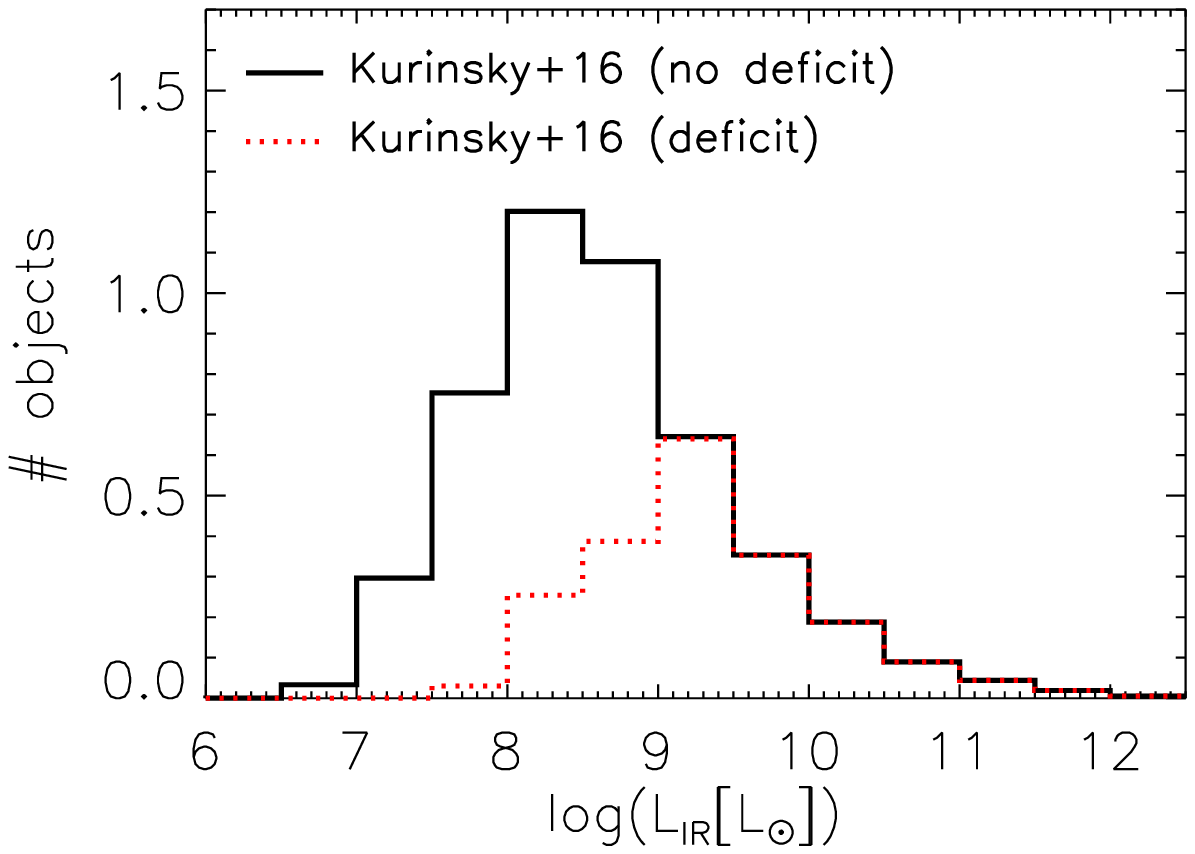}
}
  \caption{Predicted IR luminosity distributions of galaxies detectable in at least 2 lines (with 0.1\,h integration). \textit{On the left:} 
  for the \citet{Kurinsky2016} model, comparison between the results obtained using the 3 different assumptions for the low-$L$ end of the IR LFs. \textit{On the right:} for the reference model (\citealt{Kurinsky2016} model with case 2 low-$L$ end assumption), comparison between the predictions obtained using the reference correlations between L$_{\rm IR}$ and L$_{\rm PAH}$ and those obtained using the assumption of a factor 10 deficit in L$_{\rm PAH}$ for all the galaxies having $L_{\rm IR}<10^{9}\,L_{\sun}$.}
  \label{fig:LIR_distributions}
\end{figure}

\begin{figure} 
\centering
    \includegraphics[trim=0.7cm 0.0cm 3.0cm
    0.0cm,clip=true,width=0.49\textwidth, angle=0]{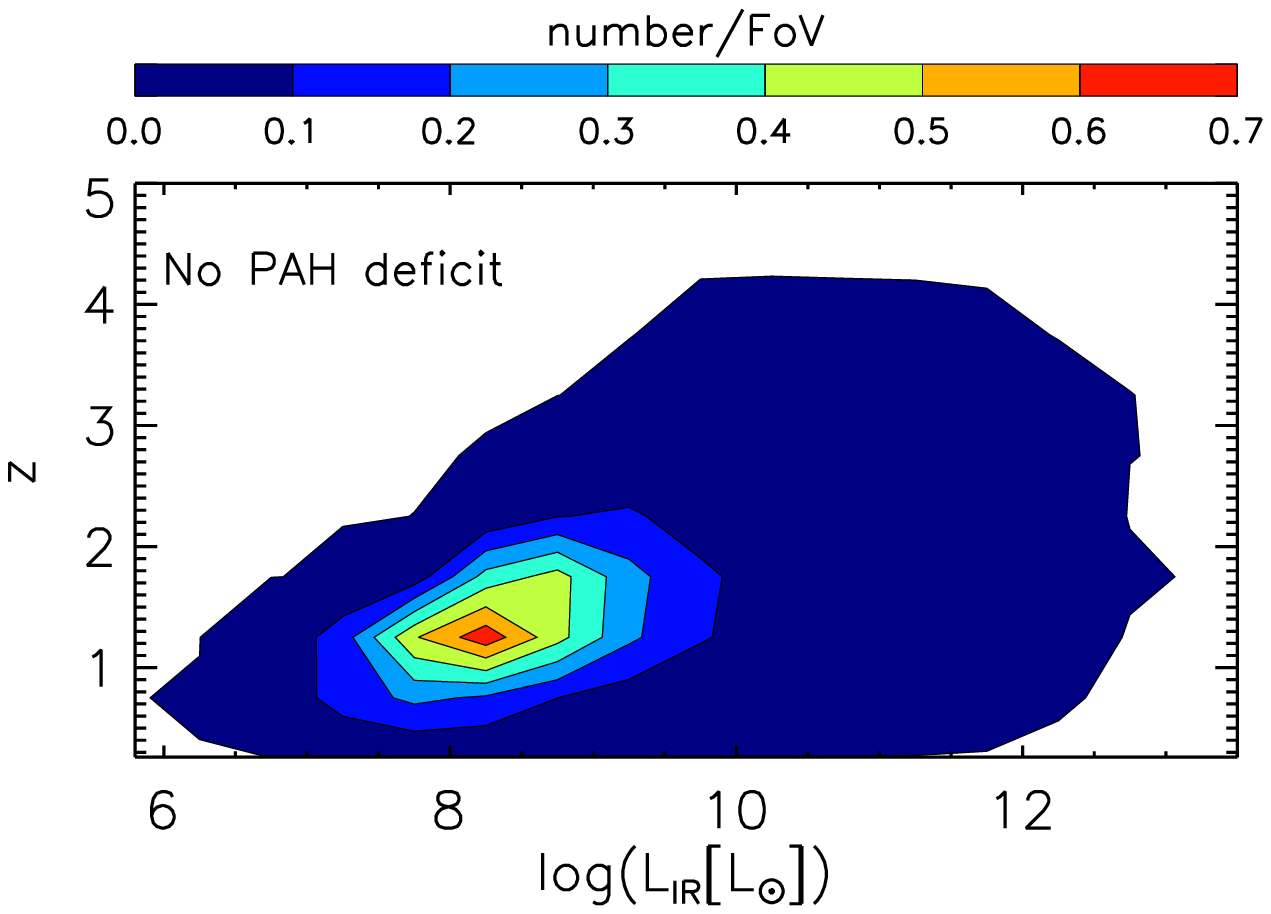}
        \includegraphics[trim=0.7cm 0.0cm 3.0cm
    0.0cm,clip=true,width=0.49\textwidth, angle=0]{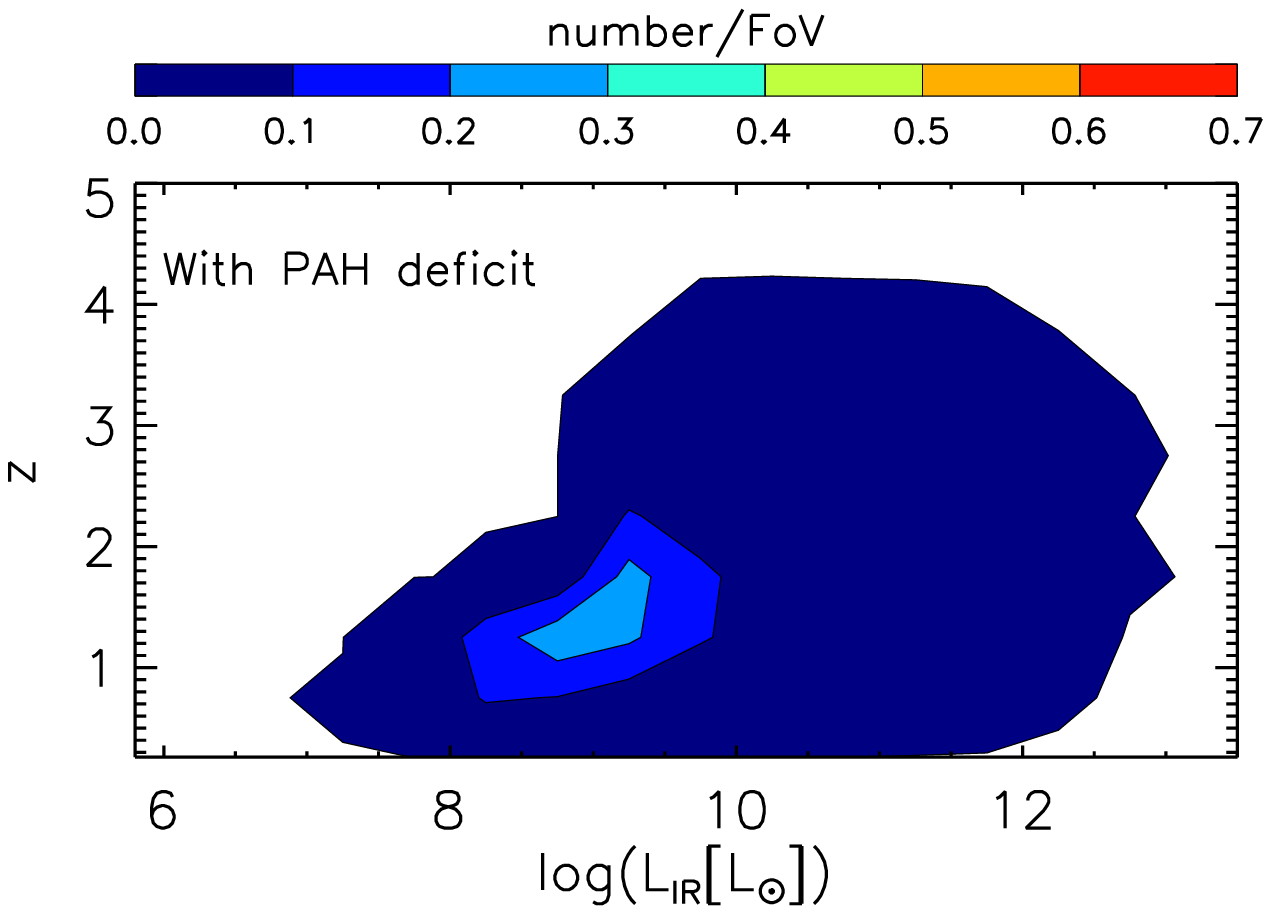}
  \caption{Two dimensional density function, on the log($L_{\rm IR}$)-z plane, of the predicted sources detectable in at least 2 lines by a MIRI MRS serendipitous survey covering the area of the FoV in 0.1\,h integration. \textit{On the left:} predictions for the reference model (\citealt{Kurinsky2016} model with case 2 low-$L$ end assumption) and the reference correlations between L$_{\rm IR}$ and L$_{\rm PAH}$. \textit{On the right:} predictions for the reference model and the assumption of a factor 10 deficit in L$_{\rm PAH}$ (for all the galaxies having $L_{\rm IR}<10^{9}\,L_{\sun}$). The different colors correspond to the number of detected sources (see the color bar on the top).}
  \label{fig:histo2D_MRS_E}
\end{figure}

\subsection{Implications for the study of high-$z$ low-$L$ galaxies \label{sec:lowl_discussion}}


Figure~\ref{fig:LIR_distributions} shows the luminosity distributions of serendipitously-detected galaxies detectable in at least 2 lines - per FoV and for an integration time of 0.1\,h. We compare these distribution for the three different assumptions about the low-luminosity end of the IR LF and including or not a PAH deficit. We can see that in all cases the peak of the detected sources is located at very low luminosities ($\lesssim$10$^{9}\,L_{\sun}$). Figure~\ref{fig:histo2D_MRS_E} shows how the log(L$_{\rm IR}$)-$z$ plane distribution of the detected (in at least 2 lines) sources varies with the PAH deficit assumption. The peak of the detectable galaxies is located at IR luminosities $\lesssim$10$^{9}\,L_{\sun}$ and at redshifts $\sim$1-2. Using the \citet{Cai13} model we obtained consistent results. Therefore all the considered scenarios suggest that, for the first time, MIRI MRS will be able to detect in a few minutes extremely faint galaxies, up to high redshifts, and with good statistics. 

Here we consider sources detectable in at least 2 lines, because robust redshift measurements will be provided for them enabling reconstruction of the low-$L$ end of the IR luminosity function. However, we verified that considering sources detectable in at least 1 or 3 lines, the peaks of the redshift and luminosity distributions are generally located in the same bins as the 2 line case. The only difference is in the number of sources detected which increases by $\sim$40-50\% in the 1 line case and decreases by $\sim$10-20\% in the 3 line case.

These galaxies will be detected essentially in PAH lines. As mentioned in Sect.~\ref{sect:intro}, the relationship between line and continuum emission in this luminosity regime is highly uncertain and only a minor fraction of the IR luminosity is probably due to SF (see Subsect.~\ref{subsect:evol}). However follow-up observations of these sources would permit us to study the evolution of their physical properties and to derive diagnostics and indicators, that are currently calibrated only for their high-$L$ counterparts. 

For example, observing these low-$L$ serendipitous sources with the JWST Near-Infrared Spectrograph\footnote{It will provide integral-field spectroscopy with a wavelength coverage of 1-5$\mu$m in medium-resolution.} (NIRSpec; \citealt{Posselt2004}; \citealt{Ferruit2012}), in medium-resolution spectroscopy, would allow us to detect them in both H$_{\alpha}$ and H$_{\beta}$ lines (for galaxies at z$\gtrsim$1). Combining NIRSpec H$_{\alpha}$/H$_{\beta}$ measurements with MRS PAH ones, for a sample of serendipitous (z$\gtrsim$1) low-$L$ galaxies, will allow us to calibrate the PAH luminosity as SFR indicator in this regime, through the extinction-corrected $H_{\alpha}$ luminosity (corrected from the Balmer decrement; ratio of $H_{\alpha}$/H$_{\beta}$). A similar approach to the method is presented in \citet{Shipley2016} for high-L star-forming galaxies.

\section{Conclusions}\label{sect:concl}

In this paper we build upon earlier works in \citet{Bonato2014a,Bonato2014b,Bonato2015} which includes relations for line to total IR luminosity for a large number of SF and AGN mid-IR spectral lines. With these, we make specific predictions for the detectability of the said lines in different redshift and IR luminosity regimes. We find that only a few minutes integration of MRS pointed observations of \textit{Herschel}-selected sources are sufficient to obtain spectroscopic redshifts for all of them and to investigate the role of SF and AGN therein through a combination of PAH equivalent widths and fine-structure line ratios. 

We use the recent model for the evolution of the IR LF from \citet{Kurinsky2016} coupled with three different assumptions for the unknown low-$L$ slope of the luminosity function, and accounting for the PAH deficit in low-$L$, low-metallicity galaxies to investigate the likely number serendipitous galaxy detections in the same pointed MRS observations described above. We find that each 0.1\,h pointed MRS observation can result in tens of serendipitous detections. Most of them ($>$83\%) are PAHs, while AGN detections will be only $<$6\% of the total.


Such serendipitous surveys will allow us, for the first time and for free, to detect very low-luminosity galaxies up to high redshifts, and with good statistics. The bulk of these sources have $\sim$10$^{8-9}\,L_{\sun}$ and $z\sim1-2$. This is a completely unexplored regime. Thus serendipitous MIRI surveys will allow us to study the properties of these intermediate redshift dwarf galaxies and to test galaxy evolution models, in particular the faint end of the IR LFs. In fact, serendipitous detections collected from about one hundred FoVs (with 0.1\,h integration) will be sufficient for a tight estimation of the slope of the low-L end.

These serendipitous observations will be able to achieve unexplored levels of SFR and BHAR. Even with short integration times of a few minutes, the improvement over \textit{Herschel} is impressive: about three orders of magnitude in SFR and one order in BHAR. This inclusion of lower luminosity sources will allow us to reconstruct the PAH line luminosity functions, continuum luminosity functions and SFR functions (with $\sim$100 FoVs in 0.1\,h integration) much more accurately than currently possible.



\section*{Acknowledgements} We thank Prof. George Rieke for helpful discussions. We are grateful to the anonymous referee for many constructive comments that helped us in improving this paper. MB is supported by NASA-ADAP13-0054. AS and JM acknowledge support through NSF AAG\#1313206. AP acknowledges support from NSF AAG \#1312418. MN has received funding from the European Unions Horizon 2020 research and innovation programme under the Marie Sklodowska-Curie grant agreement No 707601.


\bibliographystyle{apj}
\bibliography{mir}

\end{document}